\documentstyle[aps,epsf]{revtex}
\title{
Baryon magnetic moments in the effective quark Lagrangian approach
}
\author{Yu.A.Simonov$^1$, J.A.Tjon$^{2,3}$ and J.Weda$^2$ \\
$^1$State Research Center, ITEP, Moscow, Russia\\ 
$^2$Institute for theoretical Physics, University
of Utrecht, The Netherlands\\ $^3$KVI, University of
Groningen, The Netherlands}
\date{\today}
\newcommand{\be}{\begin{equation}}
\newcommand{\ee}{\end{equation}}  
\def\fun#1#2{\lower3.6pt\vbox{\baselineskip0pt\lineskip.9pt
\ialign{$\mathsurround=0pt#1\hfil
##\hfil$\crcr#2\crcr\sim\crcr}}}
\newcommand{\vex}{\mbox{\boldmath${\rm x}$}}
\newcommand{\veH}{\mbox{\boldmath${\rm H}$}}
\newcommand{\veA}{\mbox{\boldmath${\rm A}$}}

\newcommand{\vemu}{\mbox{\boldmath${\rm \mu}$}}

\newcommand{\vek}{\mbox{\boldmath${\rm k}$}}
\newcommand{\vey}{\mbox{\boldmath${\rm y}$}}
\newcommand{\ver}{\mbox{\boldmath${\rm r}$}}

\newcommand{\vesig}{\mbox{\boldmath${\rm \sigma}$}}
\newcommand{\vep}{\mbox{\boldmath${\rm p}$}}
\newcommand{\veq}{\mbox{\boldmath${\rm q}$}}

\newcommand{\ven}{\mbox{\boldmath${\rm n}$}}
\newcommand{\verho}{\mbox{\boldmath${\rm \rho}$}}

\newcommand{\veDel}{\mbox{\boldmath${\rm \Delta}$}}

\newcommand{\vetau}{\mbox{\boldmath${\rm \tau}$}}

\newcommand{\veJ}{\mbox{\boldmath${\rm J}$}}
\newcommand{\veM}{\mbox{\boldmath${\rm M}$}}
\newcommand{\veP}{\mbox{\boldmath${\rm P}$}}
\newcommand{\veQ}{\mbox{\boldmath${\rm Q}$}}
\newcommand{\veal}{\mbox{\boldmath${\rm \alpha}$}}
\newcommand{\vegam}{\mbox{\boldmath${\rm \gamma}$}}

\newcommand{\venab}{\mbox{\boldmath${\rm \nabla}$}}
\newcommand{\lan}{\langle}
\newcommand{\ran}{\rangle}

\begin{document}

\maketitle

\begin{abstract}
 An effective quark Lagrangian is derived from first principles 
through bilocal gluon field correlators. It is used to write down 
equations for baryons, containing both perturbative and nonperturbative 
fields. As a result one obtains  magnetic moments of octet and decuplet 
baryons without introduction of constituent quark masses
and using only string  tension as an input. 
Magnetic moments come out on average  in reasonable agreement with 
experiment, except for nucleons and $\Sigma^-$. The predictions for the
proton and neutron are shown to be in close agreement with the empirical 
values once we choose the string tension such to yield the proper 
nucleon mass. 
Pionic corrections to the nucleon magnetic moments
have been estimated. In particular, the total result
of the two-body current contributions are  found to be small.
Inclusion of the anomalous magnetic moment contributions from pion and
kaon loops leads to an improvement of the predictions.
\end{abstract}

\section{  Introduction}

\noindent
 The QCD dynamics of $q\bar q$ and $3q$ systems is governed
by two basic phenomena: confinement and chiral symmetry breaking
(CSB), which should be treated in a fully relativistically covariant
 way.  Confinement is usually introduced for static quarks via the
 area law of the Wilson loop \cite{1} or equivalently through the field
 correlators in the Field Correlator Method (FCM) \cite{2,3}.

 For spinless quarks, or neglecting spin--dependent mass corrections,
 one can envisage a self--consistent method which treats confinement
 as the area law also for  light quarks in a relativistically
 covariant way. Such method was introduced originally in \cite{4} for
 mesons, in \cite{5} for baryons,  and in \cite{6} for heavy--light mesons, and
 later on in  \cite{7} the method was generalized taking into
 account the dynamical degrees of freedom of the QCD string, which
 naturally appears due to the area law.

As a result Regge trajectories have been found in \cite{7} with the
correct string slope $(2\pi \sigma)^{-1}$.
It was realized later on, that the method used in \cite{4}-\cite{7}
can be more generally developed in the framework of the so-called einbein 
formalism, see \cite{7'}-\cite{9'}.
Spin corrections have been considered in \cite{8} for heavy mesons and in
\cite{6} for heavy--light ones. In the general case of light quarks spin-dependent 
correlations have been introduced in \cite{10'}, and for gluons in \cite{kaisi}. 
For a general review with explicit formulae see \cite{sir1}.
Baryon Regge trajectories have been found in \cite{5}.
In all  cases the basic formalism is the FCM and the Feynman--Schwinger
(or world-line) path integral representation \cite{3,fes,sitj1} which
 is well suited for relativistic quarks when spin is considered as a
 perturbation.

 The main difficulty which was always present in the method, was the
 perturbative treatment of spin degrees of freedom (which is
 incorrect, e.g., for the pion) and absence of spontaneous CSB effects  
in general \cite{12}.
 Recently a new type of formalism was suggested to treat
 simultaneously confinement and CSB and a nonlinear equation was
 derived for a light quark in the field of heavy antiquark \cite{13}.
 This equation derived directly from QCD Lagrangian was found to
 produce linear confinement and CSB for the light quark  and
 the explicit form of the effective quark mass operator $M(x,y)$ was
 defined obeying both these properties.

The eigenvalues and eigenfunctions  of the nonlocal and 
nonlinear equations have been determined  and a nonzero condensate
 was computed in \cite{sitj2},
 confirming that CSB is really present in the equations.
 In an additional study \cite{sitj3} it was demonstrated that magnetic
 field correlators do not contribute to the large distance confinement, 
however strongly modify the confinement for lowest levels and heavy-light 
masses corrected in this way are favourably compared in \cite{sitj3} to 
the experiment and results of other calculations.

Moreover, it was shown in \cite{sish1} that lattice data strongly
support the dominance of the Gaussian (bilocal) correlator,
estimating the correction due to higher correlators to 1-2\%.
Since the method of \cite{13} is quite general and allows to treat also
multiquark systems, it can be applied  to the $q\bar q$ and $3q$
systems, to find dynamical equations for them, which contain
confinement and CSB \cite{16'}. To make these  equations tractable, one
systematically exploits the large $N_c$ limit, and mostly confine
ourselves to the simplest field correlators -- the so--called
Gaussian approximation;  it was in particular shown in \cite{13} that 
the sum over all correlators does not change the qualitative results.
However, the kernel of equations becomes much more  complicated.

In the present paper we study the baryon magnetic moments based on the
derived effective Lagrangian without constituent quark masses. 
The paper is organized as follows.
In Section 2 the general effective quark Lagrangian from the
standard QCD  Lagrangian is obtained by integrating out gluonic
degrees of freedom, and the nonlinear equation for the single quark
propagator $S$ (attached to the string in a gauge--invariant way) is
derived, following the procedure in \cite{16'}.
Section 3 is devoted to the baryon Green's function, which can be expressed 
in the lowest order of our approximation scheme (neglecting gluon and pion 
exchanges) in terms of 3 independent quark Green's function, resulting in
a Hamiltonian as a sum of three quark terms.
In section 4 the next order approximation is written down when perturbative 
gluon exchanges are taken into account, including the nonperturbative interaction 
between quarks violating the factorized form of the zeroth order approximation.
The next section is devoted to the calculation of magnetic moments of baryons 
both in octet and decuplet representations of SU(3) flavour group. In
section 6 we discuss the corrections to magnetic moments due to 
pion exchange contributions.
 
\section{
 Effective quark Lagrangian}

\noindent
 As was discussed in the previous section, one can obtain an effective
 quark Lagrangian by averaging over background gluonic fields. We
 shall repeat this procedure  following \cite{13} now paying special
 attention to the dependence on the  contour in the  definition of
 contour gauge, and  introducing the operation of averaging over
 contour manifold.
  The QCD partition function for
 quarks and gluons can be written as
 \be
 Z=\int DAD\psi D\psi^+ {\rm exp} [L_0+L_1+L_{{\rm int}}],
\label{1}
 \ee
  where we are using Euclidean metric and define
  \be
  L_0=-\frac14\int d^4x(F^a_{\mu\nu})^2,
\label{2}
  \ee
  \be
  L_1=-i\int~^f\psi^+(x)(\hat \partial+m_f)~^f\psi(x)d^4x,\label{3}
  \ee
  \be
  L_{\rm int}=\int~^f\psi^+(x) g\hat A(x)~^f\psi(x)d^4x.
\label{4}
\ee
Here and in what follows $~^f\psi_{a\alpha}$ denotes quark operator
  with flavour $f$, color $a$ and bispinor index $\alpha$.

  To express $A_\mu(x) $ through $F_{\mu\nu}$ one can use the
  generalized Fock--Schwinger gauge \cite{15} with the contour $C(x)$ from
  the point $x$ to $x_0$, which can also lie at infinity,
  \be
  A_\mu(x)=\int_c F_{\lambda\beta} (z)
  \frac{\partial z_\beta (s,x)}{\partial x_\mu}
  \frac{\partial z_\lambda}{\partial s}
  ds.
\label{5}
  \ee
   Now one can integrate out the gluonic field $A_\mu(x)$, and
  introduce an arbitrary integration over the set of contours $C(x)$
  with the weight $D_\kappa(C)$, since $Z$ is gauge invariant it
  does not depend on the contour  $C(x)$. One obtains
  \be
  Z=\int D\kappa
 (C)D\psi D\psi^+{\rm  exp} \{L_1+L_{\rm eff}\},
\label{6}
 \ee
  where the
  effective quark Lagrangian $L_{\rm eff}$ is defined as
  \be
   {\rm
  exp} L_{\rm eff}=\langle {\rm exp} \int~^f\psi^+\hat A~^f\psi
  d^4x\rangle_A.
\label{7}
   \ee
Using the cluster expansion, $L_{\rm eff}$ can be written
as an infinite sum containing averages $\langle (\hat
A)^k\rangle_A$.  At this point one can exploit the Gaussian
approximation, neglecting all correlators $\langle (\hat
A)^k\rangle$ of degree  higher than $k=2$.  Numerical accuracy
of this approximation was discussed and tested in \cite{sish1}.
One expects that for static quarks corrections to Gaussian
approximation amount to  less than 2-3\%.

   The resulting effective Lagrangian is quartic in $\psi$,
   \be
   L^{(4)}_{\rm eff}=\frac{1}{2N_c} \int d^4x
   d^4y^f\psi^+_{a\alpha}(x)~^f\psi_{b\beta}(x)~^g\psi^+_{b\gamma}(y)
   ~^g\psi_{a\delta}(y)J_{\alpha\beta;\gamma\delta}(x,y)+O(\psi^6),
\label{8}
   \ee
   \be
   J_{\alpha\beta,\gamma\delta}(x,y)=(\gamma_{\mu})_{\alpha\beta}
   (\gamma_\nu)_{\gamma\delta} J_{\mu\nu}(x,y)
\label{9}
   \ee
   and $J_{\mu\nu}$ is expressed as
   \be
   J_{\mu\nu} (x,y)=g^2\int^x_C\frac{\partial
   u_\omega}{\partial x_\mu} du_\varepsilon \int^y_C \frac{\partial
   u_{\omega'}}{\partial  y_\nu}
   du_{\varepsilon'}\frac{tr}{N_c}\langle
   F_{\varepsilon\omega}(u)
   F_{\varepsilon'\omega'}(v)\rangle.
\label{10}
   \ee
$L_{\rm eff}$, Eq.~(\ref{8}), is written in the contour gauge \cite{15}.
   It can be identically rewritten in the
   gauge--invariant form if one substitutes parallel
   transporters $\Phi(x,x_0),\Phi(y, x_0)$ (identically
   equal to unity in this gauge) into Eqs.~(\ref{8}) and (\ref{10}),
   multiplying each $\psi(x)$ and $\psi(y)$ respectively
   and replacing $F(u)$ in Eq.~(\ref{10}) by
   $\Phi(x,u)F(u)\Phi(u,x_0)$ and similarly for $F(v)$.

   After that $L_{\rm eff}$ becomes gauge--invariant, but in
   general contour--dependent, if one keeps only the
   quartic term (\ref{8}), and neglects all higher terms. A
   similar problem occurs in the cluster expansion of
   Wilson loop, when one keeps only lowest correlators,
   leading to the (erroneous) surface dependence of the
   result.
   The situation here is the same as with a sum of QCD
   perturbation series, which depends on the
   normalization mass $\mu$ for any finite number of
   terms in the series. This unphysical dependence is
   usually treated by fixing $\mu$ at some physically
   reasonable value $\mu_0$ (of the order of the inverse
   size of the system).

   The integration over contours $D\kappa(C)$ in (\ref{6})
   resolves this difficulty in a similar way.
      Namely, the partition function $Z$ formally does not
   depend on contours (since it is integrated over a set
   of contours) but depends on the  weight
   $D\kappa(C)$. We choose this weight in such a way,
   that the contours would generate the string of
   minimal length between $q$ and $\bar q$. Thus the
   physical choice of the contour corresponds to the minimization of
   the meson (baryon) mass over the class of strings,
   in the same way as the choice of $\mu=\mu_0$
   corresponds to the minimization of the dropped higher
   perturbative terms.
   As a practical outcome, we shall keep the integral
   $D\kappa(C)$ till the end
   and finally use it to minimize the string between the quarks.

   Till this point we have made only one approximation --neglected
   all field correlators except the Gaussian one.
Recent lattice calculations (see Refs.~\cite{dig,bali})
estimate the accuracy of this approximation at the level of few percents.
 Now one must use
   another approximation, i.e. assume a large $N_c$ expansion and keep the
   lowest term. As was shown in \cite{13} this enables one to replace in
   (\ref{8}) the colorless product 
   $~^f\psi_b(x)~^g\psi_b^+(y)=
   tr (~^f\psi(x)\Phi(x, x_0)\Phi(x_0,y)~^g\psi^+(y)) $
   by  the quark Green's function
   \be
   ~^f\psi_{b\beta}(x)~^g\psi^+_{b\gamma}(y)\to
   \delta_{fg}N_cS_{\beta\gamma}(x,y).
\label{11}
   \ee
$L^{(4)}_{\rm eff}$ assumes the form
   \be
   L^{(4)}_{\rm eff}=-i\int d^4xd^4y~^f\psi^+_{a\alpha}(x)
   ~^fM_{\alpha\delta}(x,y) ~^f\psi_{a\delta}(y),
\label{12}
   \ee
   where the quark mass operator is
   \be
   ~^fM_{\alpha\delta}(x,y)=-J_{\mu\nu}(x,y) (\gamma_{\mu}~^fS(x,y)
   \gamma_\nu)_{\alpha\delta}.
\label{13}
   \ee
~From (\ref{12}) it is evident that $~^fS$ satisfies 
   \be
   (-i\hat \partial_x-im_f)~^fS(x,y)-i\int~^fM(x,z) d^4
   z~^fS(z,y)=\delta^{(4)}(x-y).
\label{14}
   \ee
   Eqs.~(\ref{13})-(\ref{14}) have been first derived in \cite{13}. From (\ref{6}) and
   (\ref{12}) one should expect that at large $N_c$ the $q\bar q$ and $3q$
   dynamics is expressed through the quark mass operator (\ref{13}), which
   should contain both confinement and CSB.
Indeed, the analysis performed in Refs.~\cite{13}-\cite{sitj3} reveals that 
confinement is present in the long--distance form of $M(x,y)$, when both
distances $|\mbox{\boldmath ${\rm  x}$}|, |\mbox{\boldmath ${\rm
y}$}|$ of light quark from heavy antiquark (placed at
$\mbox{\boldmath $ {\rm x}$} =0$) are large.

We shall  now make several simplifying assumptions, to clarify the
structure of $M(x,y)$. First of all we take the class of contours
$C$ going from any point $x=(x_4, \mbox{\boldmath $ {\rm x}$})$ to
the point $(x_4,0)$ and then to $(-\infty)$ along the $x_4$ axis.
For this class the corresponding gauge was studied in \cite{18}.
Secondly, we take the dominant part of $J_{\mu\nu}$ in (\ref{13}),
namely $J_{44}$, which is proportional to the correlator of
color--electric fields. This yields a linear confining interaction,
while the other components $J_{ik}, J_{i4}, J_{4i}, i=1,2,3$ have been
neglected, containing magnetic fields and yielding momentum dependent
corrections. (It is easy to take into account these contributions
in a more detailed analysis).

The correlator $\langle FF\rangle$ in (\ref{10}) can be expressed
through the scalar correlator $D(x), $ defined as \cite{2},
\be
\frac{trg^2}{N_c}\langle F_{\alpha\beta}(u) \Phi(u,v)
F_{\gamma\delta} (v)\Phi(v,u)\rangle =
D(u-v)(\delta_{\alpha\gamma} \delta_{\beta\delta}
-\delta_{\alpha\delta}\delta_{\beta\gamma})+O(D_1),
\label{15}
\ee
where the correlator $D_1$, not contributing to confinement, is
neglected. As a result one has for $M$ \cite{sitj2,sitj3}
\be
~^fM_{C_{x_4}}(x,y)=
~^fM^{(0)}I+~^fM^{(i)}\hat \sigma_i
+~^fM^{(4)}\gamma_4+~^fM^{(i)}_\gamma \gamma_i.
\label{16}
\ee
Here we have  defined
\be
\hat \sigma_i=\left(
\begin{array}{ll}
\sigma_i&0\\
0&\sigma_i
\end{array}\right).
\label{17}
\ee
The dominant part of $M$, $~^fM^{(0)}$ is linearly growing at large
$|\mbox{\boldmath ${\rm x}$}|, |\mbox{\boldmath ${\rm y}$}|$ and in
the most simple case of Gaussian form of $D(x)$, can be written as
\be
 ~^fM^{(0)}(x,y)=\frac{1}{2T_g\sqrt{\pi}}
e^{-\frac{(x_4-y_4)^2}{4T_g^2}}\sigma|\frac{\mbox{\boldmath ${\rm x}$}
+\mbox{\boldmath ${\rm y}$}}{2}|\tilde \delta ^{(3)}
(\mbox{\boldmath
${\rm x}$}-\mbox{\boldmath${\rm y}$})
\label{18}
 \ee  where   $T_g$ is the gluon correlation length, and
   $\tilde \delta$ is a
smeared $\delta$--function, which can be represented as \cite{sitj2,sitj3}
 \be
\tilde \delta^{(3)}
(\mbox{\boldmath
${\rm x}$}-\mbox{\boldmath${\rm y}$})
\approx exp (-\frac{|
\mbox{\boldmath
${\rm x}$}-\mbox{\boldmath${\rm y}$}
|^2}{b^2})(\frac{1}{b\sqrt{\pi}})^3,~~b\sim 2T_g.
\label{19}
\ee
Here again $T_g$ is the gluon correlation length, which enters $D(u)$ as
$D(u)= D(0) {\rm exp}(-\frac{u^2}{4T_g^2})$. We are now in the
position to derive the $q\bar q$, $ 3q$ Green's function, which will be
done in the next section. \\

\section{
    Equations for the baryon Green's function}

\noindent
    Equations for the $3q$ system can be written in the same way as
    for the $q\bar q$ system.
    We again shall assume the large $N_c$ limit in the sense, that
    $1/N_c$ corrections from $q\bar q$ pairs to the quark Green's
    function and the effective mass can be neglected. We now write
    down the explicit expressions for $N_c=3$.

    The initial and final field operators are
    \be
    \Psi_{in}(x,y,z) = e_{abc}
    \Gamma^{\alpha\beta\gamma}\psi_{a\alpha}(x,C(x))
    \psi_{b\beta}(y,C(y))\psi_{c\gamma}(z, C(z))
\label{34}
    \ee
    with the notations: $a,b,c,$ are color indices, $\alpha,\beta,
    \gamma$ are Lorentz   bispinor indices and transported quark
    operators are
    \be
    \psi_{a\alpha}(x,C(x))=(\Phi_C(x,\bar x)\psi_{\alpha}(\bar
    x))_a.
\label{35}
    \ee
The contour $C(x)$ in $\Phi_C$ can be  arbitrary , but it is
    convenient to choose  it in  the same class of contours that is
    used in $D\kappa (C)$ and  in the generalized Fock--Schwinger
    gauge \cite{15}. $\Gamma^{\alpha \beta \gamma}$ is the  Lorentz spinor
    tensor securing proper baryon quantum numbers.  One can also
    choose other operators, but it does not influence the resulting
    equations.
    In Eq.~(\ref{34}) we have omitted flavour indices in $\Gamma$ and
    $\psi(x,C)$, to be easily restored in final expressions.

    Using now the effective Lagrangian (\ref{12}) valid at large $N_c$, we
    obtain for the $3q$ Green's function.
    $$
    G^{(3q)}(x,y,z|x',y',z')=
    $$
    \be
   \frac1N\int D\kappa(C)D\psi
    D\psi^+\Psi_{\rm fin}(x',y',z')\Psi^+_{\rm in}(x,y,z)
    {\rm exp} (L_1+L_{\rm eff}).
\label{36}
     \ee
     Integrating out quark degrees of
    freedom and neglecting the determinant at large $N_c$, one has \be
    G^{(3q)}=\int D\kappa(C) (e\Gamma) (e'\Gamma')\{S(x,x') S(y,y')
    S(z,z')+ {\rm perm}\},
\label{37}
 \ee 
where for simplicity color and bispinor indices are
suppressed together with parallel transporters in initial and final states.

    One can also define unprojected (without $\Gamma,\Gamma')$ $3q$
    Green's function $G^{(3q)}_{\rm un}$ with 3 initial and 3 final
    bispinor indices instead of projected by $\Gamma, \Gamma'$
    quantum numbers of baryon.
    Assuming that minimization over contours $D\kappa(C)$ reduces to
    the single choice of the contours (the single string junction
    trajectory minimizing the mass of  baryon), one can write
    equation for $G_{\rm un}^{(3q)}$:
    $$
    (-i\hat\partial_x-im_1-i\hat M_1)
    (-i\hat\partial_y-im_2-i\hat M_2)
    (-i\hat\partial_z-im_3-i\hat M_3)
    G_{\rm un}^{(3q)}=
    $$
    \be
    \delta^{(4)}(x-x')
    \delta^{(4)}(y-y')
    \delta^{(4)}(z-z')
\label{38}
    \ee
    with e.g. $\hat M_1 G\equiv \int M(x,u) G(u,x') d^4 u$.
    One can simplify the form (\ref{37}) for $G^{(3q)}$ taking into account
    that $M(x,x')$ actually does not depend on $\frac{x_4+x'_4}{2}$.
Hence the interaction kernel of $G^{(3q)}$ does not depend on
    relative energies, as in \cite{21}. Similarly to \cite{21,20} one can
    introduce Fourier transform of $G^{(3q)}$ in time components and
    take into account energy conservation $E=E_1+E_2+E_3$. One
    obtains
 $$
    G^{(3q)}(E,E_2, E_3)\simeq \int D\kappa(C) (e\Gamma) (e'\Gamma')
   $$
   \be
   \times
    \frac{1}{(E-E_2-E_3-H_1)(E_2-H_2)(E_3-H_3)},
\label{39}    \ee
    where we have used the notation
    \be
    H_i=m_i\beta^{(i)}+
\mbox{\boldmath
    ${\rm  p}$}^{(i)}
\mbox{\boldmath
    ${\rm  \alpha}$}^{(i)}+
    \beta^{(i)}M(
\mbox{\boldmath
    ${\rm  r}$}^{(i)}-
\mbox{\boldmath
    ${\rm r} $}^{(0)}).
\label{40}
    \ee
Moreover, we have taken in $M(x,x')$ the limit of small $T_g$ and the set
    of contours in $D\kappa(C)$ passing from the point $
\mbox{\boldmath
    ${\rm  r}$}^{(i)}$
    to some (arbitrary) point $
\mbox{\boldmath
    ${\rm r}$}^{(0)}$.
    As in \cite{21} one can now integrate over $E_2, E_3$ to obtain
    finally
    \be
    G^{(3q)}(E,
\mbox{\boldmath
    ${\rm r}$}_i,
\mbox{\boldmath
    ${\rm r}$}'_i)\simeq \int D\kappa(C)
    (e\Gamma) (e'\Gamma') \frac{1}{(E-H_1-H_2-H_3)}.
\label{41}
    \ee
~From (\ref{41}) one obtains equation for the $3q$ wave function similar
    to that of $q\bar q$ system,
        \be
    (H_1+H_2+H_3-E)\psi(
\mbox{\boldmath
    ${\rm r}$}_1,
\mbox{\boldmath
    ${\rm r}$}_2,
\mbox{\boldmath
    ${\rm r}$}_3)=0,
\label{42}
    \ee
where $\mbox{\boldmath ${\rm r}$}^{(0)}$ is to be taken at the
Torricelli point.
In the nonrelativistic approximation $m_i\gg
    \sqrt{\sigma}$ one has
    \be
    \sum^3_{i=1}\left [\frac{(
\mbox{\boldmath
    ${\rm  p}$}^{(i)})^2}{2m_i}+
    \sigma|
\mbox{\boldmath
    ${\rm r}$}^{(i)}-
\mbox{\boldmath
    ${\rm  r}$}^{(0)}|\right ] \Psi= \varepsilon
    \Psi,~~ \varepsilon =E-\sum m_i
\label{43}
    \ee\\

\section{ Perturbative corrections to factorized solutions}

The effective Lagrangian (\ref{8}) and the effective mass operator $M(x,y)$,
Eq.~(\ref{13}), do not take into 
account the perturbative interaction between the quarks in the baryon.
To this  end we separate the gluonic field $A_\mu$ into a background $B_\mu$ and perturbative 
parts, $A_\mu=B_\mu+a_\mu$ and use the 't Hooft identity to integrate in the partition function 
independently over both parts of $A_\mu$ as was done in \cite{9"}. 

\begin{figure}[htb]
\epsfxsize=4.5in \epsfysize=5.5in 
\begin{center}
\hspace{0.25in} 
\epsffile{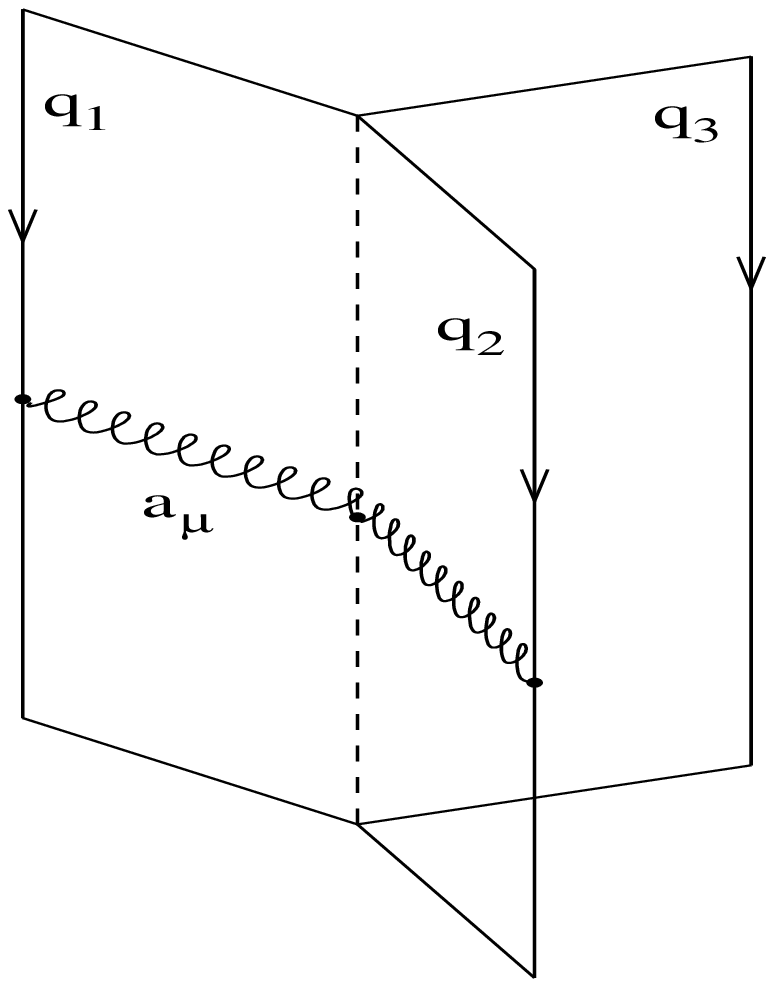}
\end{center} 
\vspace{-2in}
\caption{{\it A schematic view of the gluon propagating inside the world sheet of the string.}}
\label{fig1}
\end{figure}

We shall use the following representation of gauge transformations
\be
B_\mu\to U^+(B_\mu+\frac{i}{g}\partial_\mu) U,~~~ a_\mu\to U^+a_\mu U
\label{4.1}
\ee
 and keep for $a_\mu$ the background gauge condition \cite{dew},\cite{9"}
\be
D_\mu(B)a_\mu=0,~~~ D_\mu(B)=\partial_\mu-ig B_\mu
\label{4.2}
\ee
As a result of the perturbative gluon exchange between different quarks in the baryon 
there will appear an additional vertex in the effective Lagrangian \cite{sitj2}

\be
\Delta L=g^2\int^f\psi^+(x) \gamma_\mu^f\psi(x) \int^g\psi^+(y) 
\gamma_\nu^g\psi(y)\lan a_\mu(x) a_\nu(y)\ran dxdy.
\label{4.3}
\ee
 In what follows we shall be interested only in  the color Coulomb interaction which
results from (\ref{4.3}) assuming the simplest form of gluon propagator and neglecting at first for
simplicity the influence of the background field on it, namely 
\be
\int\lan a_\mu(x) a_\nu (y)\ran d(x_4-y_4) =\frac{\delta_{\mu\nu}C_2}{4\pi^2} 
\int \frac{d(x_4-y_4)}{(\bar x-\bar y)^2+(x_4-y_4)^2}
=\frac{\delta_{\mu\nu}C_2}{4\pi|\vex-\vey|}.
\label{4.4}
\ee
Now taking the background into account, one arrives at the picture of the gluon $a_\mu$ 
propagating inside the film -- the world sheet of the string, created by the background 
between three quark worldlines and the string junction, as is shown in 
Fig.~\ref{fig1}. 
Depending on the choice of $\ver^{(0)}$ we will get in general an
effective interaction of a 2-body or 3-body nature. 
Due to the presence of the QCD background the strength of 
the resulting Coulomb interaction is expected to be different from the 
perturbative OGE contribution and as a result different from the interaction
used for example in the Breit equation \cite{19'}.

Due to its attractive nature the color Coulomb contribution
leads to smaller baryon masses and giving rise to composite systems 
with a smaller radius. As a result the magnetic moments become smaller.
In the remaining part of the paper we neglect the effect from the Coulomb 
interaction. To study this a more involved analysis  is needed, where also
the hyperfine interaction has to be included. 

\section{Baryon magnetic moments without quark constituent masses}

Since the calculation of  magnetic moments as well as baryon masses does  
not involve large momentum transfer, one can use for that purpose the 
Hamiltonian  equation (\ref{42}). According to the results of 
section 4, $H_i$ can be represented as

\be
H_i= m_i\beta^{(i)} + \vep^{(i)} \veal^{(i)} + \beta^{(i)} M^{(i)} 
(\ver^{(i)}-\ver_0),
\label{5.1}
\ee
The baryon solution of (\ref{5.1}) can be represented as
\be
\Psi_{JM} = \Gamma^{\alpha\beta\gamma}_{JM} (f_1f_2f_3) e_{abc}
 \psi^{f_1}_{a\alpha}(\ver^{(1)}-\ver^{(0)}) 
 \psi^{f_2}_{b\beta}(\ver^{(2)}-\ver^{(0)}) 
 \psi^{f_3}_{c\gamma}(\ver^{(3)}-\ver^{(0)}),
\label{5.2}
\ee
where $a,b,c$ and $\alpha,\beta, \gamma$ 
refer to color and Lorentz indices respectively and  $f_i$ is the  flavour index.
In what follows we shall use only the lowest orbitals (lowest 
eigenvalues solutions) for quarks and therefore the orbital 
excitation indices are everywhere omitted.
The orbital wave function can be decomposed in the standard way

\be
\psi^f_\alpha(\verho)=\frac{1}{\rho} \left(
\begin{array}{l}
G(\rho)\Omega_{jlM}\\
iF(\rho)\Omega_{jl'M}
\end{array}
\right)
=
\left (
\begin{array}{l}
g(\rho)\Omega_{jlM}\\
if(\rho)\Omega_{jl'M}
\end{array}
\right),~~
\verho=\ver-\ver^{(0)}
\label{5.3}
\ee
and the color index is omitted, since the orbital satisfies a 
"white" (vacuum averaged) equation
\be
H_i\psi^{f_i}_{\alpha_i} =\varepsilon^{(i)}_{n_i}\psi^{f_i}_{\alpha_i}.
\label{5.4}
\ee
Therefore the only remnant of color is the requirement that $\Psi_{JM}$ 
be symmetric in all coordinates besides color.
~From Eq.~(\ref{42}) we see that the mass of the baryon, corresponding to
Eq.~(\ref{5.2}), is given by
\be
M_B = \sum_{n=1}^3 \epsilon_{n_i}^{(i)}
\ee
To define the magnetic moment one may introduce an external e.m. field $A$, 
$\vep^{(i)}\to \vep^{(i)}-e^{(i)}_q\veA,~~ \veA=\frac12 (\veH\times \ver),$ 
and calculate perturbatively the energy shift,
\be
\Delta E=-\vemu\veH.
\label{5.5}
\ee
Due to the symmetry of the problem, it is enough to consider only the 
perturbation of one orbital, say for the first quark,
\be
H_1\to H_1+\Delta H_1, ~~ \Delta H_1=-e^{(1)}_q\veal^{(1)}\veA.
\label{5.6}
\ee
Hence, denoting 
$\Psi^{(1)}=\left ( \begin{array}{l} \varphi^{(1)}\\
\chi^{(1)} \end{array}\right)$
$$ \lan\Delta H_1\ran=-e_q^{(1)} (\varphi^{(1)*}, \chi^{(1)*})\left ( \begin{array}{ll}
0&\vesig^{(1)} \veA\\
\vesig^{(1)}\veA &0
\end{array}\right)
\left ( \begin{array}{l}
\varphi^{(1)}\\
\chi^{(1)} \end{array}\right)=$$

\be
=-e^{(1)}_q(\varphi^{(1)*}\vesig^{(1)}\veA  \chi^{(1)*}+ \chi^{*(1)}\vesig^{(1)} 
\veA\varphi^{(1)}).
\label{5.7}
\ee
Using Eq. (\ref{5.3})  and a simple derivation given in Appendix A one 
obtains for  the contribution of the first quark to the magnetic moment 
operator in spin space
\be
\vemu^{(1)} = - \frac{2e^{(1)}_q}{3}\int g^{*}(r) f(r) r d^3 r 
\Omega^*_{jlM}\vesig^{(1)} \Omega_{jlM}.
\label{5.8}
\ee
For the lowest orbital $j=\frac12,~~ l=0, M=\frac12, \vesig \to \sigma_z,$ one obtains
\be
\mu_z \equiv 3 \mu_z^{(1)} = - 2 e_q^{(1)} \sigma^{(1)}_z 
\int g^{*}(r) f(r) rr^2 dr,
\label{5.9}
\ee
where the superscript 1 denotes the contribution of the first quark  to  
the magnetic moment. The normalization condition is
\be
\int(|g|^2+|f|^2) r^2 dr=1.
\label{5.10}
\ee
Note that everywhere we put  $\ver^{(1)}-\ver^{(0)} =\ver$.
In the case of a local linear confining interaction
using the Dirac equation one can express $ \mu^{(i)}$
through $g(r)$ only (see Appendix A for details)
\be
\mu_z^{(i)} =\frac{e^{(i)}_q\sigma^{(i)}_z}{3}\int^\infty_0\frac{|g|^2
r^2(2\sigma r+3\varepsilon)}{(\varepsilon+\sigma r)^2} dr.
\label{5.11}
\ee
Constructing the fully symmetrical $3q$ wave function for the nucleon
with total spin up one has for proton
$$
\Psi^P_{symm} = N'\left \{
 \frac23 [u_+(1) d_-(2) + d_-(1) u_+(2)] u_+(3)-\right.
$$
$$
-\frac13 [d_+(1) u_-(2) + u_-(1) d_+(2)] u_+(3)-
\frac13 [u_+(1) u_-(2) + u_-(1) u_+(2)] d_+(3)-
$$
\be
\left.
\frac13 [u_+(1) d_+(2) + d_+(1) u_+(2)] u_-(3)+
\frac23 u_+(1) u_+(2)  d_-(3)\right \},
\label{5.12}
\ee
where $N'=\frac{1}{\sqrt{2}}$, and subscripts  $(\pm)$ refer to the spin projection. 
In a similar way for the neutron one replaces $u\leftrightarrow d$ and 
obtains
$$
\Psi^n_{symm} = N'\left \{
 \frac23 [d_+(1) u_-(2) + u_-(1) d_+(2)] d_+(3)-\right.
$$
$$
-\frac13 [u_+(1) d_-(2) + d_-(1) u_+(2)] d_+(3)-
\frac13 [d_+(1) d_-(2) + d_-(1) d_+(2)] u_+(3)-
$$
\be
\left.
\frac13 [d_+(1) u_+(2) + u_+(1) d_+(2)] d_-(3)+
\frac23 d_+(1) d_+(2)  u_-(3)\right \}.
\label{5.13}
\ee
The matrix elements are computed easily
 \be
\lan \Psi^p_{symm} |e^{(1)}_q\sigma^{(1)}_z|\Psi^p_{symm} \ran =\frac13 e,
\label{5.14}
\ee
\be
\lan \Psi^n_{symm} |e^{(1)}_q\sigma^{(1)}_z|\Psi^n_{symm} \ran =-\frac29 e,
\label{5.15}
\ee
where $e$ is the charge of the proton.  From Eqs. (\ref{5.14})-(\ref{5.15}) 
one immediately gets the famous relation
\be
\frac{\mu^{(n)}}{\mu^{(p)}}=-\frac23.
\label{5.16}
\ee
Writing for identical orbitals the magnetic moment as a product
\be
\mu_B=3 \lan \Psi_{symm} |e^{(1)}_q\sigma^{(1)}_z|\Psi_{symm}\ran \lambda,
\label{5.17}
\ee
where 
\be
\lambda\equiv -\frac23 \int g^* (r)f(r) r^3 dr.
\label{5.18}
\ee
It is clear that inclusion of higher  orbitals 
will change the magnetic moment of  proton and neutron, 
similarly to the case of tritium and $^3He$, where the admixture of the orbital 
momentum $L=2$ changes the magnetic moment by 7-8\%. 
In our case the orbital momentum is brought by all 3 quarks  symmetrically, 
and these components appear in the wave function due to mixing through the tensor 
and spin-orbit forces between quarks. 

Eqs.~(\ref{5.17})-(\ref{5.18}) can readily be generalized when the quarks 
have different orbital wavefunctions. For the single quark orbitals we have
taken the solution of the Dyson-Schwinger-Dirac equation with nonlocal
kernel from Refs.~\cite{sitj2,sitj3}. Assuming for the field correlator a Gaussian
form
\be 
D(u)=D(0) exp(-u^2/4T_g^2),~ ~ D(0)=\frac{\sigma}{2\pi T_g^2}
\ee
with $T_g=0.24~fm$ the ground state orbital solution is determined. 
In Table \ref{table1} are shown the calculated ground state energy of the orbitals
for various flavour states. For the current masses we have used
$m_u=m_d=5~MeV$ and $m_s=200~MeV$.
\newpage

Using these orbitals we 
calculate the nucleon magnetic moment for various values  of the 
string tension $\sigma$. The results are also shown in Table \ref{table1}. From 
the table we see that the predictions depends sensitively on the
string tension $\sigma$. Increasing the value of $\sigma$ leads to a 
larger ground state energy of the orbitals and smaller size
of the magnetic moment. This in accordance with an
analysis, where the small component of the orbital is treated
perturbatively. Similarly the presence of a Coulomb interaction 
yields a lower ground state energy of the orbital, resulting in
a larger value in magnitude of the magnetic moment.
Close agreement with the experimental values of the magnetic moment is 
found when $\sigma=0.09~GeV^2$. In this case the mass of the 
nucleon is predicted to be $891~MeV$. It is gratifying to
see, that the magnetic moments are reasonable in the regime where 
also the predicted mass of the nucleon is close to the empirical value. 

\begin{table}[htb]
\caption{{\it 
Ground state energy $\epsilon_0$ of the orbitals 
and the predicted  magnetic moments 
of the nucleons in units of nuclear magneton for various values
of $\sigma$. 
The experimental values are also listed.}}
\label{table1}
\vspace{0.5 cm}

\begin{center}
\begin{tabular}{|ccc|cc|}
\hline
$\sigma~(GeV^2)$  &  $\epsilon_0(u,d)~(MeV)$ &  $\epsilon_0(s)~(MeV)$ 
 & $\mu_{proton}$ & $\mu_{neutron}$ \\
\hline
$0.09$	& 297 & 439	& 2.81	& -1.87 \\
$0.12$	& 342 & 482	& 2.44	& -1.63 \\
$0.15$	& 380 & 519     & 2.20	& -1.46 \\
\hline
&&experiment   		& 2.79	& -1.91 \\  
\hline
\end{tabular}
\end{center}
\end{table}

\begin{table}[htb]
\caption{{\it The magnetic moment  of the baryons in units of nuclear 
magneton for various values of $\sigma$. Calculations 
and experimental results.}}
\label{table2}
\vspace{0.5cm}

\begin{center}
\begin{tabular}{|l|cccc|}
\hline
B &  $\mu_{B}$ & $\mu_{B}$ & $\mu_{B}$ & exp \\
  &  $\sigma=0.09~GeV^2$ & $\sigma=0.12~GeV^2$ & $\sigma=0.15~GeV^2$ &  \\
\hline
p 		&   2.81 &  2.44 &  2.20 &  2.79 \\
n 		&  -1.87 & -1.63 & -1.46 & -1.91 \\
$\Sigma^{-}$	&  -1.03 & -0.89 & -0.79 & -1.16 \\
$\Sigma^{0}$	&   0.85 &  0.74 &  0.67 &   \\
$\Sigma^{+}$	&   2.72 &  2.37 &  2.14 &  2.46 \\
$\Lambda$	&  -0.66 & -0.60 & -0.56 & -0.61 \\
$\Xi^{-}$	&  -0.57 & -0.53 & -0.50 & -0.65 \\
$\Xi^{0}$	&  -1.51 & -1.34 & -1.23 & -1.25 \\
\hline
$\Delta^{++}$	&   5.62 &  4.89 &  4.39 &  4.52 \\
$\Delta^{+}$	&   2.81 &  2.44 &  2.20 &   \\
$\Delta^{0}$	&   0.00 &  0.00 &  0.00 &   \\
$\Delta^{-}$	&  -2.81 & -2.44 & -2.20 &   \\
$\Sigma^{+*}$	&   3.09 &  2.66 &  2.37 &   \\
$\Sigma^{0*}$	&   0.27 &  0.21 &  0.18 &   \\
$\Sigma^{-*}$	&  -2.54 & -2.23 & -2.02 &   \\
$\Xi^{0*}$	&   0.55 &  0.43 &  0.35 &   \\
$\Xi^{-*}$	&  -2.26 & -2.02 & -1.84 &   \\
$\Omega^{-}$	&  -1.99 & -1.80 & -1.67 & -2.02 \\
\hline
\end{tabular}
\end{center}
\end{table}
The explicit form of $\Psi_{symm}$ for other baryons are given in 
Appendix B. 
Note, that due to 
the strange quark mass their orbitals are different from those of $u,d$ quarks 
and therefore the decomposition (\ref{5.17}) has to be modified.
Some useful formulas can be found in Appendix B. 

The resulting values for baryon magnetic moments are given in Table \ref{table2}, 
where they are compared with experimental  values. 
Considering the case of $\sigma=0.12~GeV^2$ we see, that there is
a rather close agreement with the experimental magnetic moments, with the
largest deviations found for the nucleon and $\Sigma^{-}$. 
As discussed for the case of the nucleon improvement
of the predicted mass of the composite system also leads to
magnetic moments closer to the experimental values. This
applies also for the case of the $\Delta$-isobar. Hence
we may hope that the inclusion of the Coulomb and hyperfine splitting 
interaction will improve the predictions.
Moreover, pionic effects are expected to be present.
As a result significant mesonic current contributions to the
magnetic moments may occur.
In the next section we study the dominant corrections from the pion
to the one- and two-body current.

\section{Mesonic contributions}

In this section we carry out in our single orbital model an estimate of 
the magnitude of the pionic-current corrections to the magnetic moment
of the nucleon. Due to the quark-coupling to effective mesonic degrees
of freedom, one and two-body current contributions to the magnetic 
moments of the baryons exist from the virtual excitations of mesons.
Assuming as in Ref.~\cite{graz} that there exists an effective one meson exchange
between quarks in the three-quark system this leads to meson exchange
current contributions to the magnetic moment. The leading correction
is due to the pion-in-flight and pair term, see Ref.~\cite{graz}. Effects from 
the heavier mesons like the $\rho$ are in general less important. 

Our starting point is the e.m. current matrix element
\be
M_{\mu}= \left<\Psi\left| J_{\mu}(Q)\right|\Psi\right>,
\label{cur0}
\ee
where $\Psi$ is the 3-quark wavefunction and $Q$ is the photon momentum. 

We first consider the single quark current contribution.
For the single quark current operator we use
\be
J_{\mu}^{\gamma q q} \equiv 3J_{\mu}^{\gamma qq}(1)= 3 e^{(1)}_q \gamma_{\mu}^{(1)} \prod_{n=2}^3 \gamma_0^{(n)}
\label{1b}
\ee
and for the wavefunction normalization Eq.~(\ref{5.10}) for the
single particle orbitals is taken.
This choice has the nice property that the zeroth-component of the current at
$Q=0$ is found to give the proper charge of the 3-quark system, i.e.
\be
M_0=\left<\Psi\left|J_0(Q=0)\right|\Psi\right>= \sum_{n=1}^{3} e^{(n)}_q.
\ee
The result for the magnetic moment, obtained in the previous section 
can readily be recovered from our single quark current matrix element.
Following Ref.~\cite{kt}, the magnetic moment can be calculated 
by taking the curl of the space component of the current 
matrix element in the Breit system.
In doing so, the magnetic moment can be deduced from the e.m. current as
\be
\mu_{z} = \frac{e}{2M_p} G_{mag}(Q=0) = 
-\frac{i}{2} \left[ \venab_{Q} \times \veM \right]_{z}({Q=0}),
\label{5.6a}
\ee
where $M_p$ is the proton mass,  $e$ the proton charge and $G_{mag}$ is
the Sachs e.m. magnetic form factor.
The matrix element (\ref{5.6a}) can easily be evaluated in
momentum space.
Introducing the Fourier transform of the wavefunction of the single 
quark orbital
\be
\tilde{\psi}^f_\alpha(\vek)= \left (
\begin{array}{l}
\tilde{g}(k)\Omega_{jlM}\\
\tilde{f}(k)\Omega_{jl'M}
\end{array}
\right)=4\pi \int\left (
\begin{array}{l}
(-i)^{l} j_{l}(k\rho) g(\rho)\Omega_{jlM}\\
i(-i)^{l'} j_{l'}(k\rho) f(\rho)\Omega_{jl'M}
\end{array}
\right)\rho^2 d\rho,
\label{5.5a}
\ee
with $j_{l}$ the spherical Bessel functions, we
may after some algebra reduce Eq. (\ref{5.6a}) in momentum space to
\be
\mu_z=3\mu_{z}^{(1)}
=3\left<\psi_{symm}\left|e_{q}^{(1)}\sigma_{z}^{(1)}
\right|\psi_{symm}\right>\tilde{\lambda},
\label{5.8a}
\ee
We thus find,
\begin{eqnarray}
\tilde{\lambda}  = && \frac{-1}{2N}\int\int d^{3}p d^{3}q
\prod_{n=2}^{3}
\left(|\tilde{g}(k_{n})|^{2} +|\tilde{f}(k_{n})|^{2}\right)
\nonumber
\\
&& \times
\left(\tilde{g}(k_{1})\frac{4}{3k_{1}}\tilde{f}(k_{1})
-\frac{\partial\tilde{g}(k_{1})}{\partial k_{1}}
\frac{2}{3}\tilde{f}(k_{1})
+\tilde{g}(k_{1})\frac{2}{3}\frac{\partial\tilde{f}(k_{1})}
{\partial k_{1}}\right)_{Q^{2}=0},
\label{5.10a}
\end{eqnarray}
where $N$ is the normalization factor
\be
N = \int\int d^{3}p d^{3}q
\prod_{n=1}^{3} \left(|\tilde{g}(k_{n})|^{2}+
|\tilde{f}(k_{n})|^{2}\right).
\label{5.11a}
\ee
The momenta are expressed in terms of the Jacobi coordinates as,
\be
\begin{array}{lp{1cm}l}
\vek_{1}=-\frac{2}{\sqrt{3}}\veq+\frac{1}{3}\veP,  && 
\vek'_{1}=-\frac{2}{\sqrt{3}}\veq'+\frac{1}{3}\veP',  \\
\vek_{2}=\vep+\frac{1}{\sqrt{3}}\veq+\frac{1}{3}\veP, && 
\vek'_{2}=\vep'+\frac{1}{\sqrt{3}}\veq'+\frac{1}{3}\veP', \\
\vek_{3}=-\vep+\frac{1}{\sqrt{3}}\veq+\frac{1}{3}\veP, && 
\vek'_{3}=-\vep'+\frac{1}{\sqrt{3}}\veq'+\frac{1}{3}\veP'. 
\end{array}
\label{5.9a}
\ee
Imposing the Breit system, $\veP+\veP'=0$, and momentum conservation gives
$\veP'=-\veP=\veQ/2$, $\vep'=\vep$ and $\sqrt{3}(\veq-\veq')=\veQ$. 

Use has been made of the identity
\be
\left<\Omega_{jlM}(\hat{k}_{1})\left|(\hat{\vek}_{1})_{i}
(\hat{\vek}_{1})_{j}\right|\Omega_{jlM}(\hat{k}_{1})\right>
=\frac{1}{3}\delta_{ij}.
\label{5.12a}
\ee
with $l=0$ and Eqs.~(\ref{5.14}-\ref{5.15}).
The magnetic moment expression (\ref{5.17}) from the previous section
is readily  recovered when we replace the integration over the Jacobi 
momenta in Eqs.~(\ref{5.10a}-\ref{5.11a}) by $\prod_{n=1}^{3} dk_n$.

\begin{figure}
\epsfxsize=4.5in \epsfysize=5.5in 
\begin{center}
\hspace{0.25in} 
\epsffile{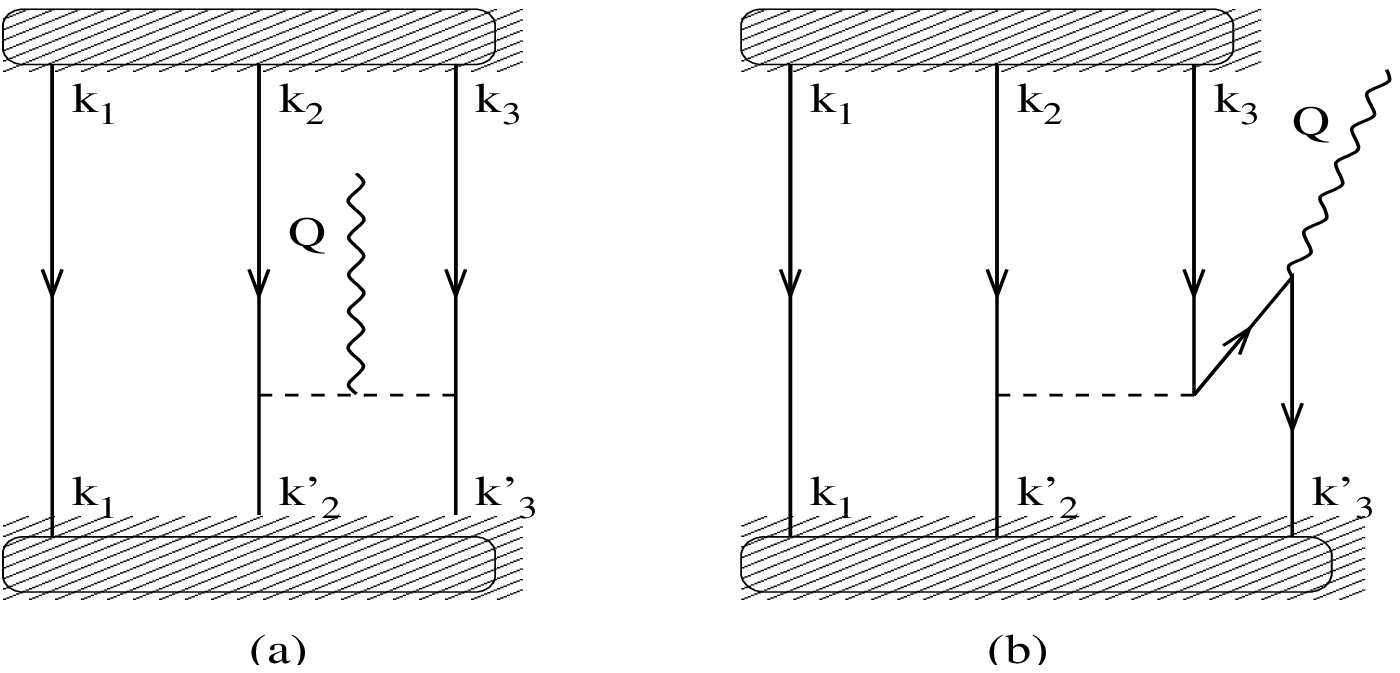}
\end{center} 
\vspace{-2in}
\caption{{\it The diagrams corresponding to the pionic contributions 
to the current: 
(a) the pion-in-flight diagram, (b) the pair term.  
The bound state of the quarks is represented by the blobs at the beginning 
and the end of the diagrams. }}
\label{fig2}
\end{figure}

We now turn to the pionic two-body current contributions,
assuming a $\gamma_5$ theory.  The resulting pion-in-flight and 
pair current operators, shown in Fig.~\ref{fig2} are given respectively by

\be
\begin{array}{c}
\veJ_{\gamma\pi\pi}^{(23)} = -2i e g_{\pi qq}^{2} 
\gamma_{5}^{(2)}\gamma_{5}^{(3)} 
\left(\vetau^{(2)}\times\vetau^{(3)}\right)_{z}
\frac{\veDel}{\left(\left(\veDel-\frac{1}{2}\veQ\right)^{2} +
m_{\pi}^{2}\right)
\left(\left(\veDel+\frac{1}{2}\veQ\right)^{2}+m_{\pi}^{2}\right)} 
\\
\frac{\Lambda_{\pi}^{4}}{\left((\veDel-\frac{1}{2}
\veQ)^{2}+\Lambda_{\pi}^{2}\right)\left((\veDel+\frac{1}{2}\veQ)^{2}+
\Lambda_{\pi}^{2}\right)}
\left(1+\frac{(\veDel-\frac{1}{2}\veQ)^{2}+m_{\pi}^{2}}{(\veDel+\frac{1}{2}\veQ)^{2}+\Lambda_{\pi}^{2}}+
\frac{(\veDel+\frac{1}{2}\veQ)^{2}+m_{\pi}^{2}}{(\veDel-\frac{1}{2}\veQ)^{2}
+\Lambda_{\pi}^{2}}\right)
\label{pion}
\end{array}
\ee
and
\be
\begin{array}{c}
\veJ_{\gamma N \bar{N}}^{(23)} = -ie g_{\pi qq}^{2}
\gamma_{5}^{(2)}\gamma_{5}^{(3)}
\left(\vetau^{(2)}\times\vetau^{(3)}\right)_{z}
\left[
\frac{\left(\gamma^0 -1\right)^{(3)}}{4m_q}\vegam^{(3)}
\frac{1}{\left(\left(\veDel-\frac{1}{2}\veQ\right)^{2} + m_{\pi}^{2}\right)}
\frac{\Lambda_{\pi}^{4}}{\left((\veDel-\frac{1}{2}\veQ)^{2}+\Lambda_{\pi}^{2}\right)^{2}}.
\right.
\\
\left.
-\frac{\left(\gamma^0 -1\right)^{(2)}}{4m_q}\vegam^{(2)}
\frac{1}{\left(\left(\veDel+\frac{1}{2}\veQ\right)^{2} + m_{\pi}^{2}\right)}
\frac{\Lambda_{\pi}^{4}}{\left((\veDel+\frac{1}{2}\veQ)^{2}+\Lambda_{\pi}^{2}\right)^{2}}
\right]
\end{array}
\label{paar}
\ee

In Eqs.~(\ref{pion})-(\ref{paar}) $Q$ is the photon momentum, 
$\Delta=\vep-\vep'$. A monopole form factor with cutoff mass 
$\Lambda_{\pi}=675~MeV$ has been used. 
The last two terms in the last factor in Eq.~(\ref{pion}) correspond to  
contact terms, which are needed to satisfy current conservation.
The quark-propagator in Eq.~(\ref{paar}) has been replaced by its negative 
energy part,
\be
\frac{i}{\not p-m}\Rightarrow
\frac{i}{2\sqrt{\vep^{2}+m^{2}}}\frac{\vep\vegam-m+\sqrt{\vep^{2}+m^{2}}\gamma^{0}}{p^{0}+\sqrt{\vep^{2}+m^{2}}}
\approx\frac{i}{4m}\left(\gamma^{0}-1\right),
\label{propagator}
\ee
as the positive energy part has already been included in the single quark 
current matrixelement~\cite{chemtob}.
Moreover, the pair contribution~(\ref{paar}) consists of 4 terms where the photon
can interact with quark 2 and 3 prior and after the pion-quark interaction.

The photo-pion vertex is described by an effective interaction Lagrangian
\be
{\cal L}_{\pi\pi\gamma} = 
-\frac{1}{2}eA_{\mu}\left(\vec{\pi}\times\partial^{\mu}\vec{\pi}\right)_{z}
+\frac{1}{2}eA_{\mu}\left(\partial^{\mu}\vec{\pi}\times\vec{\pi}\right)_{z}.
\ee

~From the 2-body operators $\veJ_{2b}$, (\ref{pion}-\ref{paar}) we
may write down
the current matrix element between the 3-quark state
\be
\veM_{2b} = 3\veM_{2b}^{(1)} = 3\frac{1}{N}\int\int d^{3}p d^{3}q
\bar{\Psi} \gamma_0^{(1)} \veJ_{2b}^{(23)} \Psi.
\label{2body}
\ee
Taking the curl of Eq.~(\ref{2body}) the magnetic moment 
can be determined.  The resulting 
expressions are given in Appendix C. As a check using the obtained magnetic 
moment operators we have determined the exchange magnetic moment
contribution to the trinucleon system. Our results agree with those
obtained by Kloet and Tjon~\cite{kt}.

\begin{table}[htb]
\caption{{\it The single quark current contribution $\mu_N^{(1)}$
 to the magnetic moment in units of nuclear magneton, together
with the two-body corrections and the anomalous correction $\delta
\mu_N^{(1)}$ arising from the pion one-loop diagrams. 
Also are shown the total combined prediction of our calculations
and the experimental results}}
\label{table3}
\vspace{0.5cm}

\begin{center}
\begin{tabular}{|l|cccccc|}
\hline
&&&&&\\
N& $\mu_{N}^{(1)}$ & $\mu_{N}^{(\pi\pi\gamma)}$ & $\mu_{N}^{(N \bar{N} \gamma)}$&
$\delta \mu_{N}^{(1)}$ & $\mu_{N}^{tot}$& exp\\
\hline
& & & $\sigma=0.09~GeV^2$ & & &\\
p		&   2.81 &  0.20 & -0.21 &  0.12 &  2.92 &  2.79\\
n 		&  -1.87 & -0.20 &  0.21 & -0.16 & -2.02 & -1.91\\
\hline
& & & $\sigma=0.12~GeV^2$ & & &\\
p		&   2.44 &  0.19 & -0.18 &  0.11 &  2.56 &  2.79\\
n 		&  -1.63 & -0.19 &  0.18 & -0.14 & -1.78 & -1.91\\
\hline
& & & $\sigma=0.15~GeV^2$ & & &\\
p		&   2.20 &  0.18 & -0.16 &  0.10 &  2.32 &  2.79\\
n 		&  -1.46 & -0.18 &  0.16 & -0.13 & -1.61 & -1.91\\
\hline
\end{tabular}
\end{center}
\end{table}

To get an estimate of the exchange 
current contributions in the 3-quark case we have used for the couplings and cut off mass
the values from Ref.~\cite{graz}. They are taken to be
$g_{qq \pi}^{2}/4\pi = 0.67$.
The results for the magnetic moments are shown in Table \ref{table3}. 
Our estimates are in strong disagreement with those obtained in
Ref.~\cite{graz}. 
The pion-in-flight contribution is substantially smaller than found in
Ref.~\cite{graz} using the chiral constituent model~\cite{graz0}.
This may be partially due to the 3-quark wavefunction used,
which has a matter radius smaller than in our case. Moreover, it
contains only nonrelativistic components.
The pair contribution is found to be comparable to the
pion-in-flight term, leading to an almost cancellation of the mesonic
current pionic contributions. 

\begin{figure}
\epsfxsize=4.in \epsfysize=5.in 
\begin{center}
\epsffile{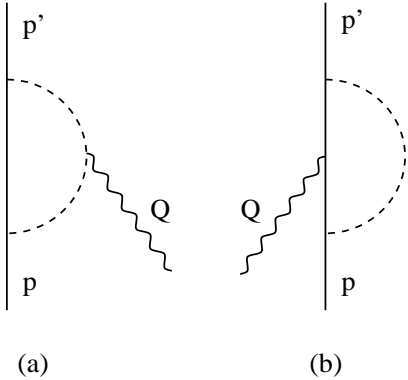}
\end{center} 
\vspace{-1.5in}
\caption{{\it The diagrams contributing to the anomalous magnetic moment
of
the single quark}}
\label{fig3}
\end{figure}

The presence of mesonic degrees of freedom will modify the single 
quark current. The resulting e.m. current operator can in
general be characterized by a large number of off-shell form 
factors~\cite{binc}-\cite{tt}, which reduces to 2 when we  
assume that the initial
and final quark is on-mass shell. Using this approximation
we may estimate the resulting anomalous magnetic $\kappa$ term due to the
the pionic contributions. Near $Q^2=0$ we have
\be
J_{\mu}^{\gamma q q} =  e_q \gamma_{\mu} +
\kappa_q \frac{ie}{2 M_p} \sigma_{\mu\nu} q_{\nu},
\label{anmu}
\ee
where $\kappa_q = \kappa_s + \kappa_v \tau_z$ for the u,d-quark.
The $\kappa$ coefficients can be determined in a simple model,
assuming that  the loop corrections are given by only the
one-loop pionic contributions to the e.m. vertex. 
Similarly as in the two-body current case we approximate the single quark 
orbital by a free quark propagation with a constituent mass given 
by the ground state orbital energy.
With the above simplifying assumptions the calculation amounts
to calculating the magnetic moment contributions of the
diagrams shown in Fig.~\ref{fig3}. 
Using the same cutoff mass regularization as for the two-body currents
we find for the anomalous magnetic moment in units of
the nuclear magneton,
\be
\kappa^{(a)}=\kappa_v^{(a)}\tau_z=
ig_{\pi qq}^2\tau_z\frac{4M_p}{3m_q^3}\int
\frac{d^4k}{\left(2\pi\right)^4}
\frac{4\left(p\cdot k\right)^2-p^2k^2}
{\left[k^2-2pk+i\epsilon\right]\left[k^2-m_{\pi}^2+i\epsilon\right]^2}
\left(\frac{\Lambda_{\pi}^2}{k^2-\Lambda_{\pi}^2}\right)^2
\left(1+2\frac{k^2-m_{\pi}^2}{k^2-\Lambda_{\pi}^2}\right)
\label{anmma}
\ee
and
\be
\kappa^{(b)}=\kappa_s^{(b)}+\kappa_v^{(b)}\tau_z=
-ig_{\pi qq}^2\frac{1-\tau_z}{2}\frac{2M_p}{3m_q^3}\int
\frac{d^4k}{\left(2\pi\right)^4}
\frac{4\left(p\cdot k\right)^2-p^2k^2}
{\left[k^2-2pk+i\epsilon\right]^2\left[k^2-m_{\pi}^2+i\epsilon\right]}
\left(\frac{\Lambda_{\pi}^2}{k^2-\Lambda_{\pi}^2}\right)^2,
\label{anmmb}
\ee
where $p$ is the momentum of the quark. For details we refer to 
Appendix D. Eq.~(\ref{anmma}) corresponds to the
coupling of the photon to the pion, Eq.~(\ref{anmmb}) to the 
coupling of the photon to the quark.

\begin{table}[htb]
\caption{{\it The quark anomalous magnetic moments 
in units of nucleon magneton 
in the one-loop approximation for various string
tension $\sigma$.The first set is the prediction for only
the pion loops, while the second set is with both pion and kaon
loops included.}}
\label{table4}
\vspace{0.5cm}

\begin{center}
\begin{tabular}{|l|ccc|}
\hline
$\sigma~ (GeV^2)$ & $\kappa_u$ & $\kappa_d$ & $\kappa_s$ \\
\hline
&pion loops&&\\
&&&\\
0.09 & 0.101 & -0.160 & 0.0\\
0.12 & 0.092 & -0.140 & 0.0 \\
0.15 & 0.085 & -0.126 & 0.0 \\
\hline
&pion  and kaon loops&&\\
&&&\\
0.09 & 0.132 & -0.151 & -0.034\\
0.12 & 0.121 & -0.133 & -0.032 \\
0.15 & 0.112 & -0.120 & -0.031 \\
\hline
\end{tabular}
\end{center}
\end{table}

In Table~\ref{table4} are shown the calculated anomalous magnetic
moments of the u, d and s quarks for $\Lambda=675~MeV$ for various
choices of $\sigma$. Clearly, the results depend on the constituent
quark masses. These are given in Table~\ref{table1} for the considered
string tensions. 

Using Eq.~(\ref{5.6a}) the $\kappa$-term in Eq.~(\ref{anmu})
yields a nucleon magnetic moment correction
\be
\delta \mu_z=3\delta \mu_{z}^{(1)}
=3 \left<\psi_{symm}\left|\kappa_q(1)\sigma_{z}(1)\right|\psi_{symm}\right>
\lambda_0
\label{kappa}
\ee
with
\be
\lambda_0 = \frac{\int r^2 dr (|g|^2-|f|^2)}{\int r^2 dr
(|g|^2+|f|^2)}
\ee
In Table~\ref{table3} the predictions for the nucleon are shown
including also the one-pion loop contributions (\ref{kappa})
and two-body currents. Our results obtained for the one-loop
corrections are smaller than reported by Glozman and
Riska~\cite{GR}. This is due to the inclusion of the lower
component in the single quark orbitals. Neglecting these we
recover the results of Ref.~\cite{GR}.
From Table~\ref{table3} we see that the proton and neutron magnetic
moment is in reasonable agreement with experiment for a string tension of
$\sigma=0.1~~GeV^2$. For this value of the string tension the model predicts
a nucleon mass of $940 MeV$, remarkably close to the empirical value.
The anomalous magnetic moment contributions are found to be
of the order of 10\%.

Due to the one-loop contributions the magnetic moments of
the other baryons are modified. Corrections from kaon loops have also 
been considered. Because of the larger kaon mass the contributions are 
expected in general to be smaller in size than those of the 
pion loops. In 
Table~\ref{table4} the calculated anomalous moment of the strange quark 
due to the kaon one-loop corrections are given.
In the calculations a cutoff mass of $\Lambda=675~MeV$ has been used.
The isoscalar and isovector anomalous magnetic moment 
pieces are also  changed by the kaon loop contributions.
From Table~\ref{table4}
we see that the kaon loop contributions are indeed smaller in
magnitude as compared to the pion loop ones.
The full results for the magnetic moments of the baryon octet and
decuplet, including the pionic exchange currents and the pion and 
kaon one-loop contributions are summarized in Table~\ref{table5}. 
For the value of the string tension $\sigma=0.1$ the overall agreement
with the experimental data is reasonable.
From the table we see that the anomalous magnetic moment contribution leads 
to an improvement of the predictions.

\newpage
\begin{table}
\caption{{\it The magnetic moment $\mu_B$
of the baryon octet and decuplet
in units of nuclear magneton, including the anomalous contribution $\delta
\mu_B$ arising from the pion and kaon one-loop diagrams and the pion
exchange corrections for different 
string tension $\sigma$. Also are shown the experimental results}}
\label{table5}
\vspace{0.5cm}

\begin{center}
\begin{tabular}{|l|lc|lc|lc|c|}
\hline
&&&&&&&\\
B &  $\delta \mu_B$ & $\mu_{B}$ & $\delta \mu_B $ &
$\mu_{B}$ & $\delta \mu_B$ & $\mu_{B}$ & exp \\
  &  $\sigma=0.09~GeV^2$ && $\sigma=0.12~GeV^2$ && $\sigma=0.15~GeV^2$ & &  \\
\hline
p		&   0.15 &  2.95 &  0.14 &  2.59 &  0.12 &  2.34 &  2.79  \\
n		&  -0.16 & -2.02 & -0.14 & -1.78 & -0.13 & -1.61 & -1.91  \\
$\Sigma^{-}$	&  -0.13 & -1.16 & -0.11 & -1.00 & -0.10 & -0.89 & -1.16 \\
$\Sigma^{0}$	&   0.00 &  0.85 &  0.00 &  0.74 &  0.00 &  0.67 &   \\
$\Sigma^{+}$	&   0.12 &  2.84 &  0.11 &  2.48 &  0.11 &  2.25 &  2.46 \\
$\Lambda$	&  -0.02 & -0.68 & -0.02 & -0.62 & -0.02 & -0.58 & -0.61 \\
$\Xi^{-}$	&   0.00 & -0.57 &  0.00 & -0.53 &  0.00 & -0.50 & -0.65 \\
$\Xi^{0}$	&  -0.06 & -1.57 & -0.05 & -1.39 & -0.05 & -1.28 & -1.25 \\
\hline
$\Delta^{++}$	&   0.26 &  5.88 &  0.24 &  5.13 &  0.22 &  4.61 &  4.52 \\
$\Delta^{+}$	&   0.07 &  2.88 &  0.07 &  2.51 &  0.07 &  2.27 &   \\
$\Delta^{0}$	&  -0.11 & -0.11 & -0.10 & -0.10 & -0.08 & -0.08 &   \\
$\Delta^{-}$	&  -0.30 & -3.11 & -0.26 & -2.70 & -0.24 & -2.44 &   \\
$\Sigma^{+*}$	&   0.15 &  3.24 &  0.14 &  2.80 &  0.13 &  2.50 &   \\
$\Sigma^{0*}$	&  -0.04 &  0.23 & -0.03 &  0.18 & -0.03 &  0.15 &   \\
$\Sigma^{-*}$	&  -0.22 & -2.76 & -0.20 & -2.43 & -0.18 & -2.20 &   \\
$\Xi^{0*}$	&   0.04 &  0.59 &  0.04 &  0.47 &  0.03 &  0.38 &   \\
$\Xi^{-*}$	&  -0.14 & -2.40 & -0.13 & -2.15 & -0.12 & -1.96 &   \\
$\Omega^{-}$	&  -0.07 & -2.06 & -0.06 & -1.86 & -0.06 & -1.73 & -2.02 \\
\end{tabular}
\end{center}
\end{table}

\section{Conclusion}

We have written down the general effective quark Lagrangian as
obtained from the standard QCD  Lagrangian by integrating 
out the gluonic degrees of freedom. Considering the baryon Green's function, 
neglecting gluon and meson exchanges, we find in lowest order of the approximation
scheme that it is given by a product of 3 independent single quark Green's functions.
As a result the Hamiltonian can be written 
as a sum of three quark terms, where
the single quark solutions satisfy the Dyson-Schwinger equation
with a nonlocal kernel.

The nonlinear equation for the single quark propagator 
$S$ (attached to the string in a gauge--invariant way) has been solved
in the Gaussian correlator approximation. 
The resulting 3-quark wavefunction has been used to 
determine the magnetic moments of the baryons.
This has been done for both the octet and decuplet of the 
SU(3) flavour group. 

Comparing the predictions we find that the magnetic moments 
are mostly in close overall agreement with the experiment for a string tension of 
$\sigma=0.1~GeV^2$. 
We find, that the predicted magnetic moment of the nucleon is
improved substantially once we choose a string tension to give a 
reasonable nucleon mass. The same applies for the $\Delta$-isobar. 
Effects due to the presence of virtual mesons
are in general expected to be important. We have estimated the pionic 
one-loop and one pion exchange contributions to the magnetic moment. 
The single quark corrections from pionic loops are found to be
of the order of 10\%, whereas the total effect of 2-body current 
contributions are predicted to be small, to be contrasted 
to the results of Ref.~\cite{graz}. This is due to the cancellation of
the pion in flight and pair term in the present model.
Because of the anomalous magnetic
contributions there seems to be somewhat an improvement of the predictions.

Our results for predictions of the magnetic moments of baryons
are encouraging, but are in need of including higher order corrections.
In particular, the mass spectrum obtained from our lowest order 
approximation does not contain the $N-\Delta$ mass splitting. This is
due to neglecting contributions like the hyperfine interaction arising from the
one gluon interaction.
It is clearly of interest to investigate how the magnetic
moments are changed when effects from color Coulomb and hyperfine
interaction are accounted for.

\section*{Acknowledgement}

This work was supported in part by the Stichting voor
Fundamenteel Onderzoek der Materie (FOM), which is sponsored by the
Nederlandse Organisatie voor Wetenschappelijk Onderzoek (NWO). 
Yu.A.~S. gratefully acknowledges the financial support by FOM and the
hospitality of the Institute for theoretical physics.
\newpage

\appendix
\section{Magnetic moment calculation in coordinate space}

The one-quark contribution to the magnetic moment can be written
as in (45)
\be
\lan \Delta H_1\ran= - e_q^{(1)} \int
(\varphi^{(1)*}\vesig^{(1)}\veA \chi^{(1)}+
\chi^{(1)*}\vesig^{(1)}\veA \varphi^{(1)}) d^3r,
\label{A.1} 
\ee
where $\veA =\frac{1}{2}(\veH \times \ver)$ is the vector potential of
external constant magnetic field.

Inserting in (A.1)
 $\varphi^{(1)}=g(r)\Omega_{jlM}$ and $\chi^{(1)} =
 if(r)\Omega_{jl'M}$, and taking into account that $\Omega_{jl'M}
 = - (\vesig\ven)\Omega_{jlM}$, one easily obtains
 \be
 \lan \Delta H_1\ran=-\frac12 e^{(1)}_q\int d^3r(g^*f+f^*g)
 r\Omega^*_{jlM}\{(\vesig\ven)(\ven\veH)-\vesig \veH\}\Omega_{jlM}
 \label{A.2}
 \ee
 Eq.(A.2) contains the matrix element $\int d\ven \Omega^*_{jlM}
 n_in_k\Omega_{jlM}$,
 which simplifies when $l=0$, so that $\lan
 n_in_k\ran=\frac13\delta_{ik}$.

In this case one obtains, taking into account relation $\lan
 \Delta H_1\ran = \Delta E= -\vemu^{(1)}\veH$,
 $$
 \vemu^{(1)} = - \frac13 e^{(1)}_q\int (g^*(r) f(r) + f^*(r)
 g(r)) r d^3 r \Omega^*_{jlM} \vesig^{(1)}\Omega_{jlM}=
$$
\be
=- \vesig^{(1)} \frac23 e^{(1)}_q\int Re(g^*(r)f(r)) r^3dr 
\label{A.3} 
\ee 
 In the case of a local scalar potential $U(r)$ 
one can further express $f(r)$ through $ g(r)$
using the Dirac equation for the one-quark state
\be
rf(r) =\frac{1}{\varepsilon +m +U(r)}
\left(\frac{d}{dr}(gr)+\frac{\kappa}{r} gr\right) \label{A.4} \ee
Introducing (A.4) into  (A.3) and integrating by parts one obtains
\be
\mu^{(i)}_z=\frac{e_q^{(i)}\sigma_z^{(i)}}{3} \int
\frac{|g|^2r^2dr}{(\varepsilon+m+U)^2}(3(\varepsilon+m+U)-rU'(r))
\label{A.5} \ee For $U(r)=\sigma r$ one obtains Eq.(49).
\newpage

\section{Magnetic moment of the multiplet}
\label{app:2}

In this Appendix the calculation of the nucleon magnetic moment is generalized
to the baryon octet and decuplet. By analogy with the fully symmetrical $3q$ 
wave function for the nucleon, Eqs.~(\ref{5.12}-\ref{5.13}), wave functions 
for the baryon multiplets can be formulated. 
The flavor octet with total spin $1/2$ up becomes,
\begin{eqnarray}
\label{ap.1}
\Psi^{p}_{symm} & = & \frac{\sqrt{2}}{6}\left\{2d_-u_+u_+ -u_-d_+u_+ -d_+u_-u_+ +2u_+d_-u_+\right. \\ && \left.-u_+u_-d_+ -u_-u_+d_+ -u_+d_+u_- -d_+u_+u_- +2u_+u_+d_-\right\}, \nonumber \\
\label{ap.2}
\Psi^{n}_{symm} & = & \frac{\sqrt{2}}{6}\left\{2u_-d_+d_+ -d_-u_+d_+ -u_+d_-d_+ +2d_+u_-d_+\right. \\ && \left.-d_+d_-u_+ -d_-d_+u_+ -d_+u_+d_- -u_+d_+d_- +2d_+d_+u_-\right\}, \nonumber \\
\label{ap.3}
\Psi^{\Sigma^+}_{symm} & = & \frac{\sqrt{2}}{6}\left\{2s_-u_+u_+ -u_-s_+u_+ -s_+u_-u_+ +2u_+s_-u_+\right. \\ && \left.-u_+u_-s_+ -u_-u_+s_+ -u_+s_+u_- -s_+u_+u_- +2u_+u_+s_-\right\}, \nonumber \\
\label{ap.4}
\Psi^{\Sigma^0}_{symm} & = & \frac{-1}{6}\left\{u_+d_-s_+ +d_+u_-s_+ +s_+d_-u_+ +s_+u_-d_+ -2u_+s_-d_+ -2d_+s_-u_+\right. \\ && \left.+u_-d_+s_+ +d_-u_+s_+ -2s_-d_+u_+ -2s_-u_+d_+ +u_-s_+d_+ +d_-s_+u_+\right. \nonumber \\ && \left.-2u_+d_+s_- -2d_+u_+s_- +s_+d_+u_- +s_+u_+d_-+u_+s_+d_- +d_+s_+u_-\right\}, \nonumber \\
\label{ap.5}
\Psi^{\Sigma^-}_{symm} & = & \frac{\sqrt{2}}{6}\left\{2s_-d_+d_+ -d_-s_+d_+ -s_+d_-d_+ +2d_+s_-d_+\right. \\ && \left.-d_+d_-s_+ -d_-d_+s_+ -d_+s_+d_- -s_+d_+d_- +2d_+d_+s_-\right\}, \nonumber \\
\label{ap.6}
\Psi^{\Lambda}_{symm} & = & \frac{\sqrt{3}}{6}\left\{u_-d_+s_+ -d_-u_+s_+ +u_-s_+d_+ -d_-s_+u_+ -u_+d_-s_+ +d_+u_-s_+\right. \\ && \left.-s_+d_-u_+ +s_+u_-d_+ +s_+d_+u_- -s_+u_+d_- -u_+s_+d_- +d_+s_+u_-\right\}, \nonumber \\
\label{ap.7}
\Psi^{\Xi^0}_{symm} & = & \frac{\sqrt{2}}{6}\left\{2u_-s_+s_+ -s_-u_+s_+ -u_+s_-s_+ +2s_+u_-s_+\right. \\ && \left.-s_+s_-u_+ -s_-s_+u_+ -s_+u_+s_- -u_+s_+s_- +2s_+s_+u_-\right\}, \nonumber \\
\label{ap.8}
\Psi^{\Xi^-}_{symm} & = & \frac{\sqrt{2}}{6}\left\{2d_-s_+s_+ -s_-d_+s_+ -d_+s_-s_+ 
+2s_+d_-s_+\right. \\ 
&& \left.-s_+s_-d_+ -s_-s_+d_+ -s_+d_+s_- -d_+s_+s_- +2s_+s_+d_-\right\}, \nonumber 
\end{eqnarray}
where the subscripts ($\pm$) refer to the spin projection.
For the flavor decuplet with total spin $3/2$ up we have
\begin{eqnarray}
\label{ap.9}
\Psi^{\Delta^{++}}_{symm}  & = & u_+u_+u_+, \\
\label{ap.10}
\Psi^{\Delta^{+}}_{symm}  & = & \frac{1}{\sqrt{3}}\left\{u_+u_+d_+ +u_+d_+u_+ +d_+u_+u_+\right\}, \\
\label{ap.11}
\Psi^{\Delta^{0}}_{symm}  & = & \frac{1}{\sqrt{3}}\left\{d_+d_+u_+ +d_+u_+d_+ +u_+d_+d_+\right\}, \\
\label{ap.12}
\Psi^{\Delta^{-}}_{symm}  & = & d_+d_+d_+, \\
\label{ap.13}
\Psi^{\Sigma^{+}}_{symm}  & = & \frac{1}{\sqrt{3}}\left\{u_+u_+s_+ +u_+s_+u_+ +s_+u_+u_+\right\}, \\
\label{ap.14}
\Psi^{\Sigma^{0}}_{symm}  & = & \frac{1}{\sqrt{6}}\left\{u_+d_+s_+ +d_+u_+s_+ +u_+s_+d_+ +s_+u_+d_+ +d_+s_+u_+ +s_+d_+u_+\right\}, \\
\label{ap.15}
\Psi^{\Sigma^{-}}_{symm}  & = & \frac{1}{\sqrt{3}}\left\{d_+d_+s_+ +d_+s_+d_+ +s_+d_+d_+\right\}, \\
\label{ap.16}
\Psi^{\Xi^{0}_{symm} } & = & \frac{1}{\sqrt{3}}\left\{s_+s_+u_+ +s_+u_+s_+ +u_+s_+s_+\right\}, \\
\label{ap.17}
\Psi^{\Xi^{-}}_{symm}  & = & \frac{1}{\sqrt{3}}\left\{s_+s_+d_+ +s_+d_+s_+ +d_+s_+s_+\right\}, \\
\label{ap.18}
\Psi^{\Omega^{-}}_{symm}  & = & s_+s_+s_+.
\end{eqnarray}
These fully symmetrical wave functions Eqs.~(\ref{ap.1})-(\ref{ap.18}) 
can be written symbolically as,
\be
\psi^{N}_{JM,symm}=\Gamma^{\alpha\beta\gamma}_{JM}(f_{1}f_{2}f_{3})\psi^{f_{1}}_{\alpha}\psi^{f_{2}}_{\beta}\psi^{f_{3}}_{\gamma}.
\label{ap.19}
\ee
As the orbital of the $s$-quark is heavier than the $u$- and $d$-quark orbitals
Eq.~(\ref{5.17}) has to be split up in contributions from the $u,d$-quark and
from the $s$-quark. Using the symmetrical wavefunction Eq.~(\ref{ap.19}) this
is realized by writing,
\be
\mu_{z}=3\mu_{z}^{(1)}=3\sum_{f_{1}f_{2}f_{3}}
\left<\Gamma^{\alpha\beta\gamma}_{JM}
(f_{1}f_{2}f_{3})\psi^{f_{1}}_{\alpha}\psi^{f_{2}}_{\beta}\psi^{f_{3}}_{\gamma}
\left|e_{q}(1)\sigma_{z}(1)\right|
\Gamma^{\alpha\beta\gamma}_{JM}(f_{1}f_{2}f_{3})\psi^{f_{1}}_{\alpha}\psi^{f_{2}}_{\beta}\psi^{f_{3}}_{\gamma}\right>
{\lambda}_{f_{1}},
\label{ap.21}
\ee
with,
\be
\label{ap.22}
{\lambda}_{f_i} = -\frac{2}{3}\int g_{f_{i}}^{*}(r)f_{f_{i}}(r)r^{3}dr.
\ee
The flavor index $f_i$ can take the values $u,d$ or $s$.
Note that $\lambda_u=\lambda_d$ as the same orbital is taken 
for the $u$- and $d$-quark.
Evaluating Eq.~(\ref{ap.21}) for the different baryon wave 
functions Eqs.~(\ref{ap.1}-\ref{ap.18}) results in the expressions in 
Table \ref{table6}.

\begin{table}[htb]
\caption{{\it The matrix elements of the e.m. current for the baryons.}}
\label{table6}
\begin{center}
\begin{tabular}{|l|c|}
\hline
N & $\mu_{N}/3$ \\
\hline
p 		& $ \frac{1}{3}{\lambda}_{u}$ \\
n 		& $-\frac{2}{9}{\lambda}_{u}$ \\
$\Sigma^{+}$	& $ \frac{8}{27}{\lambda}_{u}+\frac{1}{27}{\lambda}_{s}$ \\
$\Sigma^{0}$	& $ \frac{2}{27}{\lambda}_{u}+\frac{1}{27}{\lambda}_{s}$ \\
$\Sigma^{-}$	& $-\frac{4}{27}{\lambda}_{u}+\frac{1}{27}{\lambda}_{s}$ \\
$\Lambda$	& $-\frac{1}{9}{\lambda}_{s}$ \\
$\Xi^{0}$	& $-\frac{4}{27}{\lambda}_{s}-\frac{2}{27}{\lambda}_{u}$ \\
$\Xi^{-}$	& $-\frac{4}{27}{\lambda}_{s}+\frac{1}{27}{\lambda}_{u}$ \\
\hline
$\Delta^{++}$	& $ \frac{2}{3}{\lambda}_{u}$ \\
$\Delta^{+}$	& $ \frac{1}{3}{\lambda}_{u}$ \\
$\Delta^{0}$	& $  0$ \\
$\Delta^{-}$	& $-\frac{1}{3}{\lambda}_{u}$ \\
$\Sigma^{+*}$	& $ \frac{4}{9}{\lambda}_{u}-\frac{1}{9}{\lambda}_{s} $ \\
$\Sigma^{0*}$	& $ \frac{1}{9}{\lambda}_{u}-\frac{1}{9}{\lambda}_{s}$ \\
$\Sigma^{-*}$	& $-\frac{2}{9}{\lambda}_{u}-\frac{1}{9}{\lambda}_{s}$ \\
$\Xi^{0*}$	& $-\frac{2}{9}{\lambda}_{s}+\frac{2}{9}{\lambda}_{u}$ \\
$\Xi^{-*}$	& $-\frac{2}{9}{\lambda}_{s}-\frac{1}{9}{\lambda}_{u}$ \\
$\Omega^{-}$	& $-\frac{1}{3}{\lambda}_{s}$ \\
\hline
\end{tabular}
\end{center}
\end{table}
\newpage

\section{Pionic two-body contribution to the magnetic moment}

In this Appendix the pion-in-flight and pair contributions to the magnetic
moment of the nucleons are given.
Following Ref.~\cite{kt} these contributions are determined by taking the curl
of the pionic two-body currents 
Eq.~(\ref{2body}). The 3-quark state $\Psi$ is given by the
product of three single quark orbitals Eq.~(\ref{5.2}).
Because of symmetry considerations it suffices to calculate the magnetic 
moment contribution of pion exchange between say the second and the third quark
only and multiply the result by a factor of 3 to include the contribution of 
the other possible permutations of  quark pairs. 
 
Considering the pion-in-flight contribution first (Eq.~(\ref{pion})), 
taking the curl gives rather long expressions which can be divided into 
two parts,
\be
\delta\mu_{z}^{proton}=-\delta\mu_{z}^{neutron}=
3\left(\delta\mu_{z}^{A}+3\delta\mu_{z}^{B}\right).
\label{pion1}
\ee
The first part gives the larger contribution and can be written as,
\begin{eqnarray}
\label{pion1a}
\delta\mu_{z}^{A} & = & 
\lim_{Q^{2}\rightarrow 0} \frac{2eg_{\pi qq}^{2}}{3(2\pi)^{3}N}
\int d^{3}q d^{3}p d^{3}p' \frac{1}{(\veDel^{2}+m_{\pi}^{2})^{2}}\left(|\tilde{g}(k_1)|^{2}+|\tilde{f}(k_1)|^{2}\right)\times \\
&& \left\{\tilde{g}(k'_2)\tilde{f}(k_2)\tilde{g}(k'_3)\tilde{f}(k_3)
\frac{1}{3k_2k_3}\left(\vep\veDel-p_{z}\Delta_{z}\right)\right. \nonumber \\
&& +\tilde{g}(k'_2)\tilde{f}(k_2)\tilde{f}(k'_3)\tilde{g}(k_3)
\frac{1}{6k_2k'_3}\left((2\vep-\vep'+\sqrt{3}\veq)\veDel-(2p_{z}-p'_{z}+\sqrt{3}q_{z})\Delta_{z}\right) \nonumber \\
&& +\tilde{f}(k'_2)\tilde{g}(k_2)\tilde{g}(k'_3)\tilde{f}(k_3)
\frac{1}{6k'_2k_3}\left((2\vep-\vep'-\sqrt{3}\veq)\veDel-(2p_{z}-p'_{z}-\sqrt{3}q_{z})\Delta_{z}\right) \nonumber \\
&& \left.+ \tilde{f}(k'_2)\tilde{g}(k_2)\tilde{f}(k'_3)\tilde{g}(k_3)
\frac{-2}{3k'_2k'_3}\left(\vep'\veDel-p'_{z}\Delta_{z}\right)\right\}\times \nonumber \\
&& 
\left(1+2\frac{\veDel^{2}+m_{\pi}^{2}}{\veDel^{2}+\Lambda_{\pi}^{2}}\right)
\left(\frac{\Lambda_{\pi}^2}{\veDel^{2}+\Lambda_{\pi}^2}\right)^2. \nonumber
\end{eqnarray}
The second part comes from the curl applied to the wavefunctions 
\begin{eqnarray}
\label{pion1b}
\delta\mu_{z}^{B} & = & 
\lim_{Q^{2}\rightarrow 0} \frac{2eg_{\pi qq}^{2}}{3(2\pi)^{3}N}
\int d^{3}q d^{3}p d^{3}p' \frac{1}{(\veDel^{2}+\mu^{2})^{2}}\left(|\tilde{g}(k_1)|^{2}+|\tilde{f}(k_1)|^{2}\right)\frac{1}{3}\left(\hat{\vek}'_{2}\times\veDel\right)_{z} \\
&& \left\{
  \frac{\partial\tilde{g}(k'_2)}{\partial k'_2}\tilde{f}(k_2)\tilde{g}(k'_3)\tilde{f}(k_3)\left(\hat{\vek}_2\times\hat{\vek}_3\right)_{z}
- \frac{\partial\tilde{g}(k'_2)}{\partial k'_2}\tilde{f}(k_2)\tilde{f}(k'_3)\tilde{g}(k_3)\left(\hat{\vek}_2\times\hat{\vek}'_3\right)_{z}\right.\nonumber \\
&& - \frac{\partial\tilde{f}(k'_2)}{\partial k'_2}\tilde{g}(k_2)\tilde{g}(k'_3)\tilde{f}(k_3)\left(\hat{\vek}'_2\times\hat{\vek}_3\right)_{z}
   + \frac{\partial\tilde{f}(k'_2)}{\partial k'_2}\tilde{g}(k_2)\tilde{f}(k'_3)\tilde{g}(k_3)\left(\hat{\vek}'_2\times\hat{\vek}'_3\right)_{z} \nonumber \\
   && + \tilde{f}(k'_2)\tilde{g}(k_2)\tilde{g}(k'_3)\tilde{f}(k_3)\frac{1}{k'_2}\left(\hat{\vek}'_2\times\hat{\vek}_3\right)_{z}
\left.- \tilde{f}(k'_2)\tilde{g}(k_2)\tilde{f}(k'_3)\tilde{g}(k_3)\frac{1}{k'_2}\left(\hat{\vek}'_2\times\hat{\vek}'_3\right)_{z}\right\} \nonumber \\
&&  \times
\left(1+2\frac{\veDel^2+m_{\pi}^2}{\veDel^2+\Lambda_{\pi}^2}\right)
\left(\frac{\Lambda_{\pi}^2}{\veDel^2+\Lambda_{\pi}^2}\right)^2.
\nonumber
\end{eqnarray}
The normalization factor $N$ is the same as used before in the single quark
current contribution (Eq.~(\ref{5.11a})). 
The momenta are expressed in terms of the Jacobi coordinates Eqs.~(\ref{5.9a})
again, but from imposing the Breit system and momentum conservation we now get
$\veP'=-\veP=\veQ/2$ and $2\sqrt{3}(\veq'-\veq)=\veQ$.
In writing down these expressions use has been made of the 
spin-isospin operator sandwiched between the fully symmetric wavefunctions 
in spin-isospin and orbital space of the 3 quarks,
\be
\begin{array}{c}
\label{5.22}
\left<\psi_{symm}^{p}\left|
\left(\vetau^{(1)}\times\vetau^{(2)}\right)_{z}
\sigma^{(1)}_i\sigma^{(2)}_j
\right|\psi_{symm}^{p}\right> 
= 
-\left<\psi_{symm}^{n}\left|
\left(\vetau^{(1)}\times\vetau^{(2)}\right)_{z}
\sigma^{(1)}_i\sigma^{(2)}_j
\right|\psi_{symm}^{n}\right> 
\\
= 
-\frac{2}{3}\epsilon_{ij3}.
\end{array} 
\ee
It should be noted that the spin-isospin factor (\ref{5.22}) is
identical to that found for the tri-nucleon case.
For all the other baryon wavefunctions given in Appendix \ref{app:2} the 
matrix element of the considered two-body e.m. operators vanish, because
of the isospin structure of the e.m. operator. Hence
the considered two-body currents contribute only to the magnetic moment 
of the proton and neutron.

The second part $\delta\mu^{B}_z$ is a relativistic effect which enlarges the 
values by about 10\% and which vanishes in the static limit as is shown at
the end of this section.

In the same way the pair term can be analyzed. We find
\be
\label{paar1}
\delta\mu^{proton}_{z}=-\delta\mu^{neutron}_{z}=3\left(\delta\mu_{z}^{C}+3\delta\mu_{z}^{D}\right)
\ee
with
\begin{eqnarray}
\delta\mu_{z}^{C} & = & 
\lim_{Q^{2}\rightarrow 0}\frac{eg_{\pi qq}^{2}}{2m_{q}(2\pi)^{3}N}
\int d^{3}q d^{3}p d^{3}p' 
\frac{1}{\veDel^{2}+m_{\pi}^{2}}\frac{1}{3}
\left(|\tilde{g}(k_{1})|^{2}+|\tilde{f}(k_{1})|^{2}\right) \nonumber \\ 
&& \left\{\frac{1}{k_2}\left(\frac{1}{3}-\frac{\veDel\vek_2-\Delta_z(k_2)_z}
{\Delta^{2}+m_{\pi}^{2}}\right)\tilde{g}(k'_{2})\tilde{f}(k_{2})\tilde{g}(k'_{3})\tilde{g}(k_{3})\right. \nonumber \\
&& +\frac{1}{k'_2}\left(\frac{2}{3}+\frac{\veDel\vek'_2-\Delta_z(k'_2)_z}{\Delta^{2}+m_{\pi}^{2}}\right)\tilde{f}(k'_{2})\tilde{g}(k_{2})\tilde{g}(k'_{3})\tilde{g}(k_{3}) \nonumber \\
&& +\frac{1}{k_3}\left(\frac{1}{3}+\frac{\veDel\vek_3-\Delta_z(k_3)_z}{\Delta^{2}+m_{\pi}^{2}}\right)\tilde{g}(k'_{2})\tilde{g}(k_{2})\tilde{g}(k'_{3})\tilde{f}(k_{3}) \nonumber \\
&& \left.+\frac{1}{k'_3}\left(\frac{2}{3}-\frac{\veDel\vek'_3-\Delta_z(k'_3)_z}{\veDel^{2}+m_{\pi}^{2}}\right)\tilde{g}(k'_{2})\tilde{g}(k_{2})\tilde{f}(k'_{3})\tilde{g}(k_{3})\right\}
\left(\frac{\Lambda_{\pi}^2}{\veDel^2+\Lambda_{\pi}^2}\right)^2
\label{paar1a}
\end{eqnarray}
and
\begin{eqnarray}
\label{paar1b}
\delta\mu_{z}^{D} & = & 
\lim_{Q^{2}\rightarrow 0}\frac{eg_{\pi qq}^{2}}{6m_{q}(2\pi)^{3}N}
\int\int\int d^{3}q d^{3}p d^{3}p'
\frac{1}{\veDel^{2}+m_{\pi}^{2}}
\frac{1}{3}\left(|\tilde{g}(k_1)|^{2}+|\tilde{f}(k_1)|^{2}\right)\\
&& \left\{
  \frac{\partial\tilde{g}(k'_2)}{\partial k'_2}\tilde{g}(k_2)\left(\tilde{g}(k'_3)\tilde{f}(k_3)\left(\hat{\vek}'_2\hat{\vek}_3-(\hat{k}'_2)_{z}(\hat{k}_3)_{z}\right)
- \tilde{f}(k'_3)\tilde{g}(k_3)\left(\hat{\vek}'_2\hat{\vek}'_3-(\hat{k}'_2)_{z}(\hat{k}'_3)_{z}\right)\right)\right.\nonumber \\
&&+\tilde{g}(k'_3)\tilde{g}(k_3)\left( - \frac{\partial\tilde{g}(k'_2)}{\partial k'_2}\tilde{f}(k_2)\left(\hat{\vek}'_2\hat{\vek}_2-(\hat{k}'_2)_{z}(\hat{k}_2)_{z}\right)
   + \frac{\partial\tilde{f}(k'_2)}{\partial k'_2}\tilde{g}(k_2)\left(\hat{\vek}'_2\hat{\vek}'_2-(\hat{k}'_2)_{z}(\hat{k}'_2)_{z}\right)\right) \nonumber \\
&& - \left.\tilde{f}(k'_2)\tilde{g}(k_2)\tilde{g}(k'_3)\tilde{g}(k_3)\frac{1}{k'_2}\left(\hat{\vek}'_2\hat{\vek}'_2-(\hat{k}'_2)_{z}(\hat{k}'_2)_{z}\right)\right\}
\left(\frac{\Lambda_{\pi}^2}{\veDel^2+\Lambda_{\pi}^2}\right)^2.
\nonumber
\end{eqnarray}

In the non-relativistic limit the lower component of the wavefunction
can be expressed in the upper component as
\be
\tilde{f}(k)=-\frac{|k|}{2m_{q}}\tilde{g}(k),
\label{nrlimit}
\ee
where $m_{q}$ is the constituent mass of the quark.
As a result, the pionic two-body current contributions 
Eqs.~(\ref{pion1}-\ref{pion1b}) and Eqs.~(\ref{paar1}-\ref{paar1b})
can be simplified considerably. We obtain for the pion-in-flight
contribution
\be
\begin{array}{c}
\delta\mu^{A}_{z}=
\frac{eg_{\pi qq}^{2}}{6m_{q}^{2}(2\pi)^{3}N}
\int d^{3}q d^{3}p d^{3}p'
\left(|\tilde{g}(k_{1})|^{2}+|\tilde{f}(k_{1})|^{2}\right)
\tilde{g}(k'_{2})\tilde{g}(k_{2})\tilde{g}(k'_{3})\tilde{g}(k_{3})
\\
\frac{\veDel^{2}-\Delta_{z}\Delta_{z}}{\left(\veDel^{2}+m_{\pi}^{2}\right)^{2}}
\left(1+2\frac{\veDel^{2}+m_{\pi}^{2}}{\veDel^{2}+\Lambda_{\pi}^{2}}\right).
\left(\frac{\Lambda_{\pi}^{2}}{\veDel^{2}+\Lambda_{\pi}^{2}}\right)^2.
\label{nrmmpion}
\end{array}
\ee
For the pair term we get
\be
\begin{array}{c}
\delta\mu^{C}_{z}=
\frac{eg_{\pi qq}^{2}}{6m_{q}^{2}(2\pi)^{3}N}
\int d^{3}q d^{3}p d^{3}p'
\left(|\tilde{g}(k_{1})|^{2}+|\tilde{f}(k_{1})|^{2}\right)
\tilde{g}(k'_{2})\tilde{g}(k_{2})\tilde{g}(k'_{3})\tilde{g}(k_{3})
\\
\left(\frac{\veDel^2-\Delta_z^2}{\left(\veDel^{2}+m_{\pi}^{2}\right)^2}-\frac{1}{\veDel^2+m_{\pi}^2}\right) 
\left(\frac{\Lambda_{\pi}^2}{\veDel^2+\Lambda_{\pi}^2}\right)^2,
\label{nrmmpaar}
\end{array}
\ee
while $\delta\mu^{B}_{z}$ and $\delta\mu^{D}_{z}$ vanish.
These expressions agree with the results of Refs.~\cite{kt} and \cite{chemtob}.
\newpage

\section{Anomalous magnetic moment contributions from pion loops}

Our starting point is the e.m. currents, corresponding to the
one-loop diagrams shown in Fig.~\ref{fig3},
\begin{eqnarray}
J_{\mu}^{(a)} & = & -2ig_{\pi qq}^{2}e\tau_z\int 
\frac{d^4k}{\left(2\pi\right)^4}
\frac{\gamma_5\left(\not p-\not k+m_q\right)\gamma_5\left(2k_{\mu}+Q_{\mu}\right)}
{\left[(p-k)^2-m_q^2+i\epsilon\right]\left[k^2-m_{\pi}^2+i\epsilon\right]\left[\left(k+Q\right)^2-m_{\pi}^2+i\epsilon\right]}
\nonumber
\\
&& 
\frac{\Lambda_{\pi}^2}{k^2-\Lambda_{\pi}^2}
\frac{\Lambda_{\pi}^2}{\left(k+Q\right)^2-\Lambda_{\pi}^2}
\left(1+\frac{k^2-m_{\pi}^2}{\left(k+Q\right)^2-\Lambda_{\pi}^2}+
\frac{\left(k+Q\right)^2-m_{\pi}^2}{k^2-\Lambda_{\pi}^2}\right)
\label{7.3}
\end{eqnarray}
and
\be
J_{\mu}^{(b)}=-ig_{\pi qq}^{2}e\frac{1-\tau_z}{2}\int
\frac{d^4k}{\left(2\pi\right)^4}.
\frac{\gamma_5\left(\not p'-\not k+m_q\right)\gamma_{\mu}\left(\not p-\not k+m_q\right)\gamma_5}
{\left[(p'-k)^2-m_q^2+i\epsilon\right]\left[(p-k)^2-m_q^2+i\epsilon\right]\left[k^2-m_{\pi}^2+i\epsilon\right]}
\left(\frac{\Lambda_{\pi}^2}{k^2-\Lambda_{\pi}^2}\right)^2
\label{7.4}
\ee
Since we have assumed a finite form factor at the $\pi q q$ vertex, 
similar as in the two-body current case, the two additional terms 
are needed
in the last factor of Eq.~(\ref{7.3}) to satisfy current conservation.
>From these currents the anomalous magnetic moment has to be extracted.
By applying the Gordon decomposition to the current Eq.~(\ref{anmu}) near
$Q^2=0$ it can be seen that the anomalous magnetic moment $\kappa$ is the 
term proportional to $- \frac{e}{2M} K_{\mu}$ with
$K_{\mu}=p_{\mu}+p'_{\mu}$. 
To isolate this term the currents
are rewritten by explicit evaluation of the $\gamma$-matrix
algebra and taking the limit $Q^2\to 0$. Using
the approximation that the initial and final quark is on-mass shell
we obtain
\begin{eqnarray}
J_{\mu}^{(a)} & = & -2ig_{\pi qq}^{2}e\tau_{z}\gamma^{\nu}\int 
\frac{d^4k}{\left(2\pi\right)^4}
\frac{2k_{\mu}k_{\nu}}
{\left[k^2-2pk+i\epsilon\right]\left[k^2-m_{\pi}^2+i\epsilon\right]^2}
\left(\frac{\Lambda_{\pi}^2}{k^2-\Lambda_{\pi}^2}\right)^2 
\left(1+2\frac{k^2-m_{\pi}^2}{k^2-\Lambda_{\pi}^2}\right)
\nonumber \\
& \equiv & -2ig_{\pi qq}^2e\tau_{z}\gamma^{\nu}C_{\mu\nu}^{(a)},
\label{7.5}
\end{eqnarray}
and
\begin{eqnarray}
J_{\mu}^{(b)} & = & ig_{\pi qq}^{2}e\frac{1-\tau_z}{2}\gamma^{\nu}\int
\frac{d^4k}{\left(2\pi\right)^4}
\frac{2k_{\mu}k_{\nu}-k^2g_{\mu\nu}}
{\left[k^2-2pk+i\epsilon\right]^2\left[k^2-m_{\pi}^2+i\epsilon\right]}
\left(\frac{\Lambda_{\pi}^2}{k^2-\Lambda_{\pi}^2}\right)^2
\nonumber \\
& \equiv & ig_{\pi qq}^{2}e\frac{1-\tau_z}{2}\gamma^{\nu}C_{\mu\nu}^{(b)}.
\label{7.6}
\end{eqnarray}
As the tensors $C^{\mu\nu}$ depend only on the initial and final momenta 
they can be written as,
\be
C_{\mu\nu}^{(i)}=A_1^{(i)}K_{\mu}K_{\nu}+A_2^{(i)}K_{\mu}Q_{\nu}+A_3^{(i)}Q_{\mu}K_{\nu}+A_4^{(i)}Q_{\mu}Q_{\nu}+A_5^{(i)}g_{\mu\nu},
\label{7.7}
\ee
where $A_n^{(i)}$ are Lorentz invariants. It can readily be
seen, that only the first term $A_1^{(i)}$ contributes to the
magnetic moment. Substituting Eq.~(\ref{7.7}) in 
Eqs.~(\ref{7.5}-\ref{7.6}) and
taking the initial and final quark on-mass shell we find
for the anomalous magnetic moment corrections
\be
\kappa^{(a)}=8iM_pm_qg_{\pi qq}^2\tau_zA_1^{(a)},
\label{7.9}
\ee
\be
\kappa^{(b)}=-4iM_pm_qg_{\pi qq}^2\frac{1-\tau_z}{2}A_1^{(b)}.
\label{7.10}
\ee
The Lorentz invariant expression $A_1^{(i)}$ can immediately be
determined from the tensor $C_{\mu\nu}^{(i)}$. We get
\be
A_1^{(i)}=\frac{1}{3K^4}\left(4K^{\mu}K^{\nu}-K^2g^{\mu\nu}\right)C_{\mu\nu}^{(i)}.
\label{7.8}
\ee
Inserting Eq.~(\ref{7.8}) in (\ref{7.9}-\ref{7.10}) the expressions 
(\ref{anmma}-\ref{anmmb}) are obtained.

The kaon one-loop diagrams can be calculated in similar way. The starting 
point is the expressions Eqs.~(\ref{7.3}-\ref{7.4}) again,
where 
the mass of the pion is replaced by the mass of the kaon and the
isospin structure is changed to $(\tau_z+3 Y)/2$ and 
$-(\frac{2}{9}+\frac{4}{3}Y)$
respectively in Eqs.~(\ref{7.3}-\ref{7.4}) with $Y$ the hypercharge.
The expressions for the anomalous magnetic moment due to the kaon loop become,
\be
\kappa^{(a)}=
ig_{K qq}^2\left(\tau_z+3Y\right)\frac{2M_p}{3m_q^3}\int
\frac{d^4k}{\left(2\pi\right)^4}
\frac{4\left(p\cdot k\right)^2-p^2k^2+3m_q\left(M_q-m_q\right)p\cdot k}
{\left[k^2-2pk+m_q^2-M_q^2+i\epsilon\right]
\left[k^2-m_{K}^2+i\epsilon\right]^2}
\left(\frac{\Lambda_{K}^2}{k^2-\Lambda_{K}^2}\right)^2
\left(1+2\frac{k^2-m_{K}^2}{k^2-\Lambda_{K}^2}\right)
\label{anmmakaon}
\ee
and
\be
\kappa^{(b)}=
ig_{K qq}^2\left(\frac{2}{9}+\frac{4}{3}Y\right)\frac{2M_p}{3m_q^3}\int
\frac{d^4k}{\left(2\pi\right)^4}
\frac{4\left(p\cdot k\right)^2-p^2k^2+3m_q\left(M_q-m_q\right)p\cdot k}
{\left[k^2-2pk+m_q^2-M_q^2+i\epsilon\right]^2
\left[k^2-m_{K}^2+i\epsilon\right]}
\left(\frac{\Lambda_{K}^2}{k^2-\Lambda_{K}^2}\right)^2
\label{anmmbkaon}
\ee
with $M_q$ the mass of the intermediate quark, $m_q$ the mass of the initial
and final quark.
The coupling constant $g_{K qq}$ and the cutoff $\Lambda_{K}$ are
taken the same as for the pion loop.

\end{document}